\documentclass[reprint, superscriptaddress, amsmath, amssymb, aps, prb, floatfix]{revtex4-2}

\usepackage{comment}

\usepackage[caption=false]{subfig} 

\usepackage{txfonts}
\usepackage{stmaryrd}
\usepackage{rotating}
\usepackage{latexsym}
\usepackage{wasysym}
\usepackage[pdfencoding=auto, psdextra]{hyperref}
\hypersetup{
	colorlinks,%
	citecolor=blue,%
	filecolor=blue,%
	linkcolor=blue,%
	urlcolor=blue
}

\usepackage{mwe}
\usepackage{braket}
\usepackage{graphicx}
\usepackage{dcolumn}
\usepackage{bm}
\usepackage{float}
\usepackage{amsmath}
\usepackage{amsfonts}
\usepackage{amssymb}
\usepackage[countmax]{subfloat}
\usepackage{dcolumn} 
\usepackage{color}
\usepackage{xcolor}

\begin{document}

\preprint{APS/123-QED}

\title{Nonlocality of local Andreev conductances as a probe for topological Majorana wires}

\author{Rodrigo A. Dourado}
\affiliation{%
 \textit{Instituto de F\'isica de S\~ao Carlos, Universidade de S\~ao Paulo, 13560-970 S\~ao Carlos, S\~ao Paulo, Brazil}}

\author{Poliana H. Penteado}
\affiliation{%
 \textit{Instituto de F\'isica de S\~ao Carlos, Universidade de S\~ao Paulo, 13560-970 S\~ao Carlos, S\~ao Paulo, Brazil}}%
 \affiliation{\textit{Department of Physics, University of Basel, Klingelbergstrasse 82, CH-4056 Basel, Switzerland}}

\author{J. Carlos Egues}
\affiliation{%
 \textit{Instituto de F\'isica de S\~ao Carlos, Universidade de S\~ao Paulo, 13560-970 S\~ao Carlos, S\~ao Paulo, Brazil}}%
\affiliation{\textit{Department of Physics, University of Basel, Klingelbergstrasse 82, CH-4056 Basel, Switzerland}}

\date{\today}

\begin{abstract}
We propose a protocol based only on local conductance measurements for distinguishing trivial from topological phases in realistic three-terminal 
superconducting nanowires coupled to 
normal leads, capable of hosting Majorana zero modes (MZMs). 
By using Green functions and the scattering matrix approach, 
we calculate
the conductance matrix and the local density of states (LDOS) as functions of the asymmetry in the couplings to the  
left ($\Gamma_L$) and right ($\Gamma_R$) leads.  In the trivial phase, we find that the zero-bias local conductances are distinctively 
affected by variations in $\Gamma_R$ (for fixed $\Gamma_L$): while $G_{LL}$ is mostly constant, $G_{RR}$ decays exponentially  
as $\Gamma_R$ is decreased. In the topological phase, surprisingly, $G_{LL}$ 
and $G_{RR}$ are both 
suppressed with $G_{LL} \sim G_{RR}$. This \textit{nonlocal} suppression of $G_{LL}$ 
with $\Gamma_R$ scales with the MZM 
hybridization energy $\varepsilon_m$ and arises from the emergence of a dip in the LDOS near zero energy at the left 
end of the wire, which affects the local Andreev reflection. We further exploit this nonlocality of the local Andreev processes and the gate-controlled suppression of the LDOS by proposing a Majorana-based transistor. Our results hold for zero and low electron temperatures $T<20$ mK. For $T = 30, 40$ mK, $G_{LL}$ and $G_{RR}$ become less correlated. As an additional  nonlocal fingerprint of the topological phase at higher $T$'s, we predict modulations in our \textit{asymmetric} conductance deviation 
$\delta G^{asym}_{LL}= G_{LL}^{\Gamma_R = \Gamma_L} - G_{LL}^{\Gamma_R \ll \Gamma_L}$ that remains commensurate with the Majorana oscillations in $\varepsilon_m$ over the range $30<T< 150~\rm{mK}$.
%
\end{abstract}

\maketitle

\section{Introduction}

Semiconducting nanowires with proximity-induced superconductivity, in principle capable of hosting Majorana zero modes (MZMs)~\cite{kitaev2}, have become paradigmatic systems to investigate topological matter. Following theoretical predictions~\cite{nanowireProposalDasSarma,majoranaProposalFelix}, some of the early experiments~\cite{mourik2012,Das12,Deng12} relied on observing zero-bias (local) conductance peaks as primary signatures of MZMs. However, it was soon realized that these zero-bias peaks (ZBPs) could also arise from other mechanisms, e.g., the Kondo effect, disorder, and Andreev bound states~\cite{Franz, Prada2020}. Despite recent advances in materials and device optimization, there is still no clear-cut experimental evidence of MZMs in these systems. 

A step towards changing the above scenario was taken by Microsoft Quantum, which proposed ~\cite{nayakprotocol} and implemented~\cite{nayakExperiment} a protocol to identify topological phases in hybrid semiconductor-superconductor three-terminal devices. This protocol relies on the observation of coexisting ZBPs in the local conductance on the left $G_{LL}$ and right $G_{RR}$ ends of the nanowire and a closing and re-opening of the bulk transport gap as probed via the nonlocal conductance $G_{LR}$~\cite{rosdahl2018andreev}. Interestingly, Refs.~\cite{NonLocalLoss, Hess2022} show that trivial mechanisms can mimic similar features of the topological gap protocol.

\begin{figure} [htb!]
	\centering
	\includegraphics[width=0.41\textwidth]{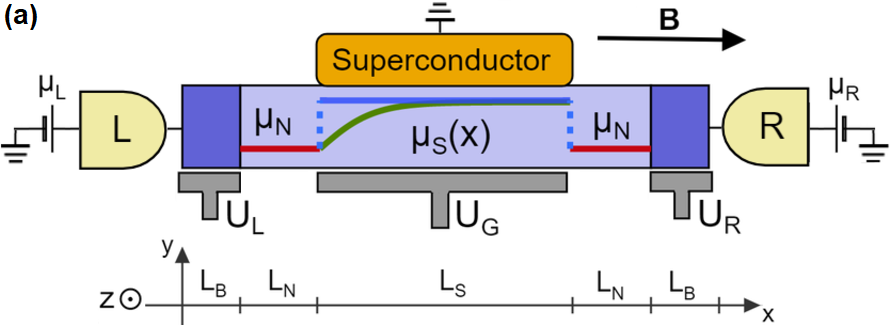}
	\quad
	\centering
	\includegraphics[width=0.45\textwidth]{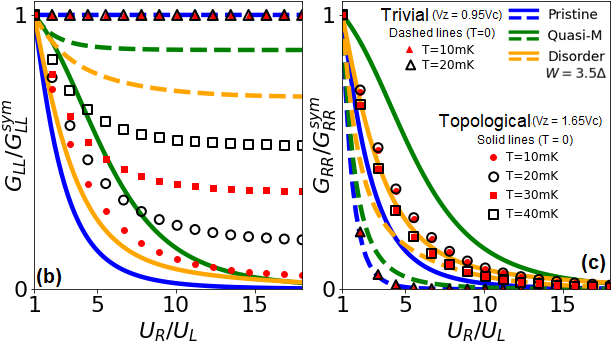}
	\caption{ (a) Semiconductor-superconductor Rashba nanowire coupled to left (L) and right (R) normal leads via gates $U_L$ and $U_R$.
The chemical potentials $\mu_S(x)$ and $\mu_N$ refer to the superconducting and normal parts of the nanowire, respectively. 
Zero-bias conductances (b) $G_{LL}$ and (c) $G_{RR}$ as functions of $U_R/U_L$. Solid (dashed) lines denote topological (trivial) phases 
for $T=0$. Symbols illustrate electron temperatures 
$T=10$, $20$, $30$, $40$ mK for topological 
($\medbullet$,$\medcirc$,$\blacksquare$,$\square$) and trivial ($\blacktriangle$,$\vartriangle$: only $10$, $20$ mK shown) phases.  
We consider pristine $\mu_S(x)=\mu$], quasi-Majorana  
[nonuniform $\mu_S(x)$] and disordered nanowires. 
In the topological regime and $T\le 20$ mK, $G_{LL} \sim G_{RR}$, and are similarly  
suppressed as $U_R/U_L$ increases, while for $T=30$, $40$ mK they become less correlated.
In the trivial regime, $G_{LL}$ remains essentially constant, while $G_{RR}$ 
is exponentially suppressed. The contrasting behavior of $G_{LL}$ and $G_{RR}$ in the topological and trivial phases should 
enable the distinction between these phases. 
Parameters used [$\rm{meV}$]: $t = \hbar^2/2m^*a^2 = 102$, 
$\mu = 1$, $\mu_N = 0.2$, $\Delta = 0.5$, $\alpha = \alpha_R/2a = 3.5$, with $a= 5~\rm{nm}$~\cite{NonLocalLoss}, and 
$U_L = 5$. Lengths [$\rm{\mu m}$]: 
$L_S = 2$, $L_N = 0.1$, and $L_B = 0.02$. }
	
	\label{fig1}
\end{figure}

In this work, we consider state-of-the-art three-terminal semiconductor-superconductor wires, Fig.~\ref{fig1}, similar to those in Refs.~\cite{rosdahl2018andreev, NonLocalLoss, nayakprotocol,  nayakExperiment,Suhas2022}. Unlike previous works~\cite{rosdahl2018andreev, nayakprotocol, nayakExperiment}, however, here we propose only the zero-bias local conductances $G_{LL}$ \textit{and} $G_{RR}$ as sufficient probes to tell apart trivial and topological phases  (bulk criterion)~\footnote{Here we are interested in zero-bias peaks. For a discussion on finite bias voltage effects see (\ref{FiniteBiasSec})}. Our proposal exploits a peculiar nonlocality of dominant \textit{local} Andreev reflection (LAR)  processes in the topological phase; comparatively, direct tunneling and crossed Andreev reflection (CAR) are less relevant in the parameter range investigated, see Appendix \ref{S1} for details [e.g.,  Fig. \ref{spectraAndCoefficients}(b)]. These retroreflections involve an incoming electron and an outgoing hole with either opposite (usual LAR) or same (spin-selective LAR~\cite{PLeeSESAR}) spins, Fig.~\ref{fig2}(a). We use the scattering matrix and Green function approaches to obtain the conductance matrix and the LDOS for generic asymmetric couplings $\Gamma_L $ and $\Gamma_R$, tunable via the barriers $U_L$ and $U_R$, respectively, Fig.~\ref{fig1}(a). 

Next, we discuss our main results for pristine and disordered/nonuniform nanowires at $T=0$, as well as at finite temperatures.

\subsection{Pristine nanowires} 

For single-subband nanowires with no disorder or nonuniformities, interestingly, as we increase the ratio $U_R/U_L$ we find that  (i)  $G_{LL} \sim G_{RR}$ decreases in the topological phase [solid blue lines in
Figs.~\ref{fig1} (b) and (c)] while (ii) in the trivial phase $G_{LL}$ is constant and $G_{RR}$ is exponentially suppressed 
[dashed blue lines in Figs.~\ref{fig1} (b) and (c)].

The \textit{nonlocal} behavior of $G_{LL}$~\footnote{
We note that nonlocal effects manifested in $G_{LL}$ were first investigated in Refs.~\cite{dassarma2013electrical, Lobos2015}  in connection with the thermal conductance, focusing on slight variations $\delta \gamma_R/\gamma_R \ll 1$ in the very asymmetric regime $\gamma_L \gg \gamma_R$.
In this perturbative limit, $\delta G_{LL} = G_{LL}(\gamma_R+ \delta \gamma_R, \gamma_L)-G_{LL}(\gamma_R, \gamma_L)$ \{this is Eq. (6) of Ref.~\cite{dassarma2013electrical}\} contains (to lowest order in $\delta \gamma_R$) only direct tunneling and CAR contributions. In our work, however, we consider a nonperturbative regime, in which $G_{LL}$ is dominated by LAR terms. We stress that $\delta G_{LL} $ in Eq. (6) of Ref.~\cite{dassarma2013electrical} \textit{is not} the same as our conductance deviation $\delta G^{asym}_{LL}$;  their couplings $\gamma_{L(R)}$ correspond to our $\Gamma_{L(R)}$} in the topological regime arises from the suppression of LAR processes at the \textit{left} end of the wire, Fig.~\ref{fig2}(a). This follows from the LDOS developing a zero-energy nonlocal dip at this same left end as the \textit{right} barrier $U_R$ is raised, Fig.~\ref{fig3}(a) [note the corresponding suppression in $G_{LL}$, inset Fig.~\ref{fig3}(b)]. This potential height increase also suppresses $G_{RR}$, even though the LDOS at the right side is enhanced. Here the spin-selective LAR~\cite{PLeeSESAR}, whose reflection probability is $A^\alpha_{\sigma \sigma}$ ($\alpha=L,R$), is the dominant process, Fig.~\ref{fig2}(b). In fact, in the pristine case,  all LAR probabilities $A^\alpha_{\sigma \sigma'}$ are mediated by (quasi-) zero-energy modes with $A_L = A_R$, where $A_\alpha = \sum_{\sigma, \sigma'} A^\alpha_{\sigma \sigma'} $. In the trivial regime, on the other hand, the usual LAR $A^\alpha_{\sigma \bar{\sigma}}$ with opposite spins $\sigma$, $\bar{\sigma}$ is the only significant contribution~\cite{setiawan2015conductance}. In this case, $A^L_{\sigma \bar{\sigma}}$ remains constant while $A^R_{\sigma \bar{\sigma}}$  exponentially decreases as $U_R$ increases ($U_L$ is kept constant). 

As a way of exploiting this nonlocality of the LAR processes and the gate-tunable suppression of the LDOS around zero energy, we propose a Majorana-based transistor, consisting of a quantum dot (QD) connected to leads $1$ and $2$, and side-coupled to our nanowire, Fig. \ref{FigTransistor}(a). A third lead  $R$ coupled to the wire controls the current through the QD. As the coupling $\Gamma_R$ (controlled by $U_R$) increases, the Majorana mode in the nanowire leaks into the dot and a zero-bias peak emerges in the LDOS, Figs.~\ref{FigTransistor}(b) and (c), thus enabling a current flow.

\subsection{Disordered/nonuniform nanowires} 

We also investigate the effects of disorder and nonuniform chemical potential profiles  (``quasi-Majoranas''), Fig.~\ref{fig1}. Quite surprisingly, even in these regimes $G_{LL}$ and $G_{RR}$ behave like the pristine case as a function of $U_R/U_L$, see Figs.~\ref{fig1} (b) and (c). Moreover, we have verified that the conductances shown in Figs.~\ref{fig1} and \ref{fig2} are stable against variations of, e.g., the applied magnetic field $\mathbf{B}$  [see Figs.~\ref{fig2}(b) and (c)], the wire carrier density, the barrier transparency, and the wire length (not shown). Hence we contend that by (independently) measuring only $G_{LL}$ and $G_{RR}$ versus $U_R/U_L$, one can distinguish the trivial ($G_{LL}\neq G_{RR}$) and topological ($G_{LL}\sim G_{RR}$) phases. In passing, we note that for true MZMs (i.e., with exactly zero energy) $G_{LL} = G_{RR}$
displays a plateau at $2e^2/h$ as a function of $U_R/U_L$. All our results are symmetric with respect to exchanging $U_R$ by $U_L$ and vice-versa. 

\subsection{Temperature effects} 

The above results for $G_{LL}$ and $G_{RR}$ hold at $T=0$ and  electron temperatures $T< 20$ mK, Figs.~\ref{fig1}(b), (c) ($\medbullet$,$\medcirc$,$\blacktriangle$,$\vartriangle$) and \ref{fig5}(a). Even though these temperatures are smaller than the experimentally reported values in similar setups~\cite{nayakExperiment,dvir2023realization, van2023electrostatic}, electron temperatures $\sim 10$ mK have been measured in related transport experiments~\cite{lowT}. For $T=30$, 40 mK, $G_{LL}$ and $G_{RR}$ become increasingly less correlated in the topological phase, see ($\blacksquare$,$\square$) in Figs.~\ref{fig1}(b), (c).   On the other hand, the \textit{asymmetric} conductance deviation $\delta G^{asym}_{LL} = G_{LL}^{U_R = U_L} - G_{LL}^{U_R \gg U_L}$ can still be used as a nonlocal probe at much higher and currently available electron temperatures. Remarkably, for  $30 <T<150$ mK, $\delta G^{asym}_{LL}$ exhibits sizable modulations that are commensurate with Majorana oscillations (in the hybridization energy $\varepsilon_m$) as the magnetic field is varied, Fig \ref{fig5}(b).  
These modulations in $\delta G^{asym}_{LL}$ provide yet another signature of the topological phase.

\section{Model Hamiltonian}

In the Nambu basis  $\Psi (x) = \{ \psi_\uparrow (x), \psi_\downarrow (x), \psi_\downarrow^\dagger (x),  -\psi_\uparrow^\dagger  (x) \}^T$, with $\psi_\sigma (x)$ the electron field operator for spin $\sigma$ at position $x$, the nanowire is modeled by~\cite{nanowireProposalDasSarma, majoranaProposalFelix}

\begin{equation} \label{Hcontinuum}
\begin{split}
    H_{\mathrm{NW}} = & \frac{1}{2} \sum_{\sigma} \int dx \Psi^\dagger (x) \left[ \left(\frac{\hbar^2 \partial_x^2}{2m^*} -i \alpha_R \partial_x \sigma_y -\mu_S (x) \right) \tau_z \right.\\
    &\left. \phantom{\frac{\hbar^2 \partial_x^2}{2m^*}}+ V_z \sigma_x + \Delta \tau_x \right] \Psi (x),
\end{split}
\end{equation}
where $m^*$ is the effective mass of the electron, $\alpha_R$ the Rashba spin-orbit coefficient, $\mu_S$ the chemical potential, $V_z$ the Zeeman energy, $\Delta$ the proximity-induced superconducting gap, and  $\sigma_i$ and $\tau_i$ the Pauli matrices acting on the spin and particle-hole spaces, respectively. 

Our nanowire also contains two outer barriers ($L_B$, $\mu_B$) followed by normal regions ($L_N$, $\mu_N$), with $\mu_B = \mu_N - U_{L, R}$, and $\Delta = V_z = 0$. The central region, described by Hamiltonian~\eqref{Hcontinuum}, has lentgh $L_S$, Fig.~\ref{fig1}(a). 
The system is coupled to $L$ and $R$ metallic leads at chemical potentials $\mu_L$ and $\mu_R$, respectively, with respect to the grounded  SC.  
We consider three different scenarios: (i) $\mu_S (x) = \mu$, uniform  (``pristine wire''), (ii) $\mu_S (x) = \mu_N + (\mu - \mu_N) \tanh[(x - L_s)/\lambda] $, a ``confining potential'' on the left side of the wire  such that $\lambda = 0.3~\mu$m controls the smoothness of the transition $\mu_N \to \mu$. 
Near-zero energy states called quasi-Majoranas emerge in the trivial phase, according to the bulk criterion $V_c = \sqrt{\mu^2 + \Delta^2}$, confined in the potential region~\cite{reproducingAkhmerov, stanescu2019robust}, (iii) $\mu_S (x) = \mu - V_{dis}(x)$, Anderson-type on-site disorder potential $V_{dis} (x)$ with values randomly taken from the interval $[-W, W]$. 
We use (i) exact numerical diagonalization to obtain the spectrum and wave functions of the Bogoliubov de-Gennes Hamiltonian, and (ii) the package KWANT~\cite{kwant} to calculate the Andreev probabilities $A^\alpha_{\sigma \sigma'}$ and the conductance coefficients $G_{\alpha \beta}$.

\section{Nonlocality of the Local Conductance}

In this section, we investigate in detail several aspects of the transmission coefficients (LAR processes) and local conductances for the Majorana nanowire depicted in Fig. \ref{fig1}(a).

\subsection{Topological Andreev bound states}

In the trivial phase, and in particular for $V_z << m^*\alpha_R^2/\hbar^2$, electrons are reflected as holes of opposite spin with probability $A^\alpha_{\uparrow \downarrow}$~\cite{setiawan2015conductance}. In contrast, the topological phase ($V_z > V_c$) is characterized by the emergence of same-spin LAR processes $A^\alpha_{\uparrow\uparrow}$ and $A^\alpha_{\downarrow\downarrow}$, in addition to opposite-spin LAR $A^\alpha_{\uparrow \downarrow}$
. Here, we will refer to these modes as topological Andreev bound states (ABS), because they arise from the hybridization of true MZMs. The eigenenergies of these quasi-degenerate states are entirely due to this hybridization~\footnote{\label{footnote1} A low-energy effective model~\cite{splittingBeenakker,nayak2008non} describing the coupling between Majoranas is represented by $H=i\varepsilon_m\gamma_1\gamma_2$, where $\gamma_i$ is the Majorana operator. In this case, the topological ABSs eigenenergies are $\pm \varepsilon_m$.}, as opposed to ordinary ABSs, whose energies are mostly due to confinement. The probabilities $A^\alpha_{\uparrow \uparrow}$ and $A^\alpha_{\downarrow \downarrow}$ are in principle measurable via polarized leads~\cite{PLeeSESAR} or spin-selective QDs~\cite{wang2022singlet}.

\begin{figure}[t]
	\centering
	\includegraphics[width=0.235\textwidth]{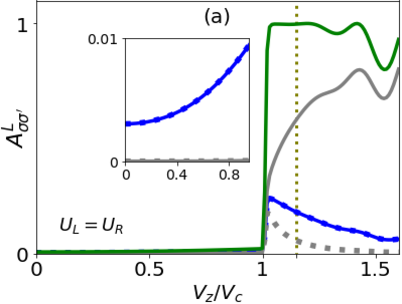}
	\hfill
	\centering
	\includegraphics[width=0.235\textwidth]{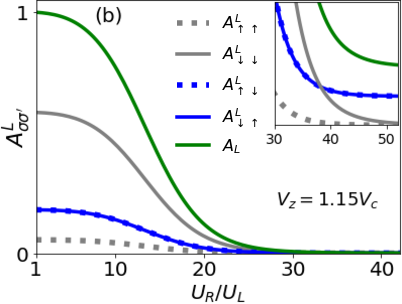}
	\quad
	\centering
	\includegraphics[width=0.235\textwidth]{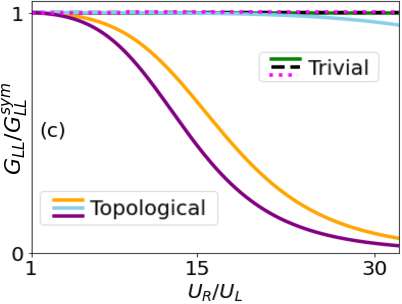}
	\hfill
	\centering
	\includegraphics[width=0.235\textwidth]{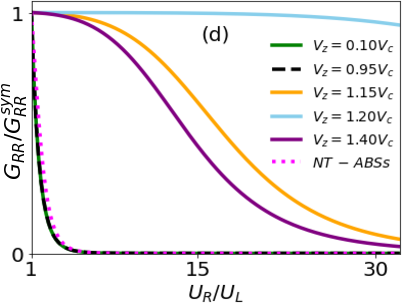}
	\caption{Local Andreev reflection probabilities $A^L_{\sigma \sigma'}$ and local conductances $G_{LL}$ and $G_{RR}$ in the pristine wire. (a) $A^L_{\sigma \sigma'}$ as a function of the Zeeman field. The topological phase is followed by the rapid increase of $A^L_{\uparrow \uparrow}$ and $A^L_{\downarrow \downarrow}$ in grey. (b) $A^L_{\sigma \sigma'}$ as a function of $U_R/U_L$ in the topological phase, $V_z = 1.15 V_c$. We observe a suppression of the LAR components as $U_R/U_L$ increases. (c) $G_{LL}$ and (d) $G_{RR}$ as  functions of $U_R/U_L$. For $V_z > V_c$ (orange, cyan, and purple curves), $G_{LL}$ = $G_{RR}$ is suppressed as  $U_R$ increases. For $V_z < V_c$ (green and black) and for fine-tuned ABSs in a nontopological (NT) wire (magenta), i.e.,  $\alpha = V_z=0$ in the superconducting region of the wire, $G_{LL}$ remains constant while $G_{RR}$ is exponentially suppressed. Parameters: same as those in Fig.~\ref{fig1}, except for $L_s = 2.5 \mu$m. In the NT case, $V_z = 1.16 V_c$ and $\mu_N = 2.58$. In (a) $U_R = 5$ meV.}
	\label{fig2}
\end{figure}

In Fig.~\ref{fig2}(a), we show $A_{\sigma \sigma'}^L$ for $U_L=U_R$ as functions of $V_z$ for opposite-spin LAR (blue curves) and same-spin LAR (grey curves); a similar plot holds for $A_{\sigma \sigma'}^R$. 
 The total LAR probability $A_\alpha$ ($A_\alpha=\sum_{\sigma \sigma'} A^\alpha_{\sigma \sigma'}$) is equal on both ends of the pristine wire, $A_L = A_R$. To see that the dominant same-spin LAR processes are directly related to topological ABSs (i.e., energy-split MZM), let us consider the strictly $E=0$ case (true Majoranas). For simplicity, we consider the $L$ lead and the SC as forming an NS junction, whose LAR matrix in zeroth order in the Rashba coupling ($2m^*\alpha_R^2/\hbar^2<<\sqrt{V_z^2-\Delta^2}$) and for $V_z>V_c$ is~\cite{PLeeSESAR}
\begin{equation} \label{eq:rhe}
    r^{he}(V_z) = 
    \begin{pmatrix} r^{he}_{\uparrow \uparrow} & r^{he}_{\uparrow \downarrow} \\ r^{he}_{\downarrow \uparrow} & r^{he}_{\downarrow \downarrow} \end{pmatrix} = \begin{pmatrix}\frac{V_z-\sqrt{V_z^2-\Delta^2}}{2V_z} & -\frac{\Delta}{2V_z} \\ -\frac{\Delta}{2V_z} &\frac{V_z+\sqrt{V_z^2-\Delta^2}}{2V_z}
     \end{pmatrix}.
\end{equation}
Within this approximation, $A^L_{\downarrow \downarrow} \approx |r^{he}_{\downarrow \downarrow}|^2\approx1-\Delta^2/2V_z^2$, $A^L_{\uparrow \uparrow} \approx |r^{he}_{\uparrow \uparrow}|^2 \approx \Delta^4/16V_z^4$, $A^L_{\uparrow \downarrow} = A^L_{\downarrow \uparrow}\approx |r^{he}_{\uparrow \downarrow}|^2 \approx \Delta^2/4V^2_z$, and $A_L\approx 1$ (up to $O(\Delta/V_z)^2$), which qualitatively describe the behavior shown in Fig.~\ref{fig2}(a). Note that $A^L_{\downarrow \downarrow}$ oscillates as $V_z$ increases, exhibiting a dip at $V_z \approx 1.5 V_c$; at this point, direct tunneling and CAR processes are favored \cite{splittingBeenakker}, see Appendix \ref{lowenergyH}. This oscillating behavior follows from the lowest topological ABS pair crossing at $E=0$ as $V_z$ varies (Majorana oscillations)~\cite{Prada2012, DasSarmaSplitting, Rainis2013}. In an NSN junction, $A^R_{\uparrow \uparrow} \approx1-\Delta^2/2V_z^2$, $A^R_{\downarrow \downarrow} \approx \Delta^4/16V_z^4$, $A^R_{\uparrow \downarrow} = A^R_{\downarrow \uparrow}\approx \Delta^2/4V^2_z$, such that $A_R \approx 1$. Note that as $V_z$ increases, the dominat component is $A^R_{\uparrow \uparrow}$. Here the MZM on the right side of the wire couples more strongly to the spin-up component of the incident electron, as opposed to the MZM on the left side, which couples more strongly to the spin-down, see Appendix \ref{lowenergyH}. In this approximate description of the pristine wire, we verify that $A_L=A_R$.

\subsection{Nonlocality of local Andreev reflection}

In the topological phase, the LAR processes on both sides of the wire are sensitive to the increase of $U_R/U_L$, as shown in Fig.~\ref{fig2}(b), in which we choose $V_z = 1.15 V_c$ [dotted vertical line in Fig.~\ref{fig2}(a)], $U_L = 5$, and vary $U_R$. As the right barrier $U_R$ increases, the left LAR probability $A^L_{\sigma \sigma'}$ decreases. Similarly, the right LAR probability $A^R_{\sigma \sigma'}$ is nonlocal for variations of the left barrier $U_L$. 

At zero bias and $T=0$, the local conductance reads 
$G_{\alpha \alpha } =\frac{2e^2}{h} (2 A_{\alpha} + T_{\bar{\alpha} \alpha}+A_{\bar{\alpha} \alpha})$, where 
$T_{\bar{\alpha} \alpha}$ and $A_{\bar{\alpha} \alpha}$ are the probabilities of an incoming electron in lead 
$\alpha=L \ (R)$ to be transmitted as an electron or as a hole, respectively, to lead $\bar{\alpha}=R \ (L)$~\cite{flensbergBCSCharge}. 
In Figs.~\ref{fig2}(c) and (d), we show, respectively, $G_{LL}$ and $G_{RR}$ as functions of $U_R/U_L$ in the topological phase for 
$V_z = 1.15 V_c$ (orange curve), $V_z = 1.20 V_c$ (cyan), and $V_z = 1.40 V_c$ (purple). We observe that they are both equally 
suppressed following the nonlocal behavior of $A_L=A_R$ in Fig.~\ref{fig2}(b). The hybridization $\varepsilon_m$ 
between the MZMs on the left and right sides of the wire is responsible for this nonlocality. Hence the $G_{LL} \sim G_{RR}$ 
suppression is more prominent for $V_z = 1.15 V_c$ and $V_z = 1.40 V_c$ as compared to $V_z = 1.20 V_c$ since 
$\varepsilon_m (V_z = 1.40 V_c) > \varepsilon_m(V_z = 1.15 V_c) >> \varepsilon_m(V_z = 1.20 V_c)$, where $\varepsilon_m$ is the 
energy of the lowest topological ABS mode. For the cyan curve, we choose $V_z$ near a parity crossing point, such that $\varepsilon_m$ 
is close to zero (see, e.g., Fig.~\ref{spectraAndCoefficients} in Appendix \ref{S1}). In this case, the dependence on $U_R$ is 
noticeable only for $U_R/U_L \sim 30$. In the limit $\varepsilon_m \to 0$, we regain $G_{LL}=G_{RR}=2 e^2/h$, i.e., the 
quantized strictly zero-energy Majorana conductance \cite{PLeeSESAR}.

In the trivial phase, $V_z = 0.10 V_c$ (green curve) and $V_z = 0.95 V_c$ (dashed black line), $G_{LL}$ does not have any dependence on $U_R$, while $G_{RR}$ exponentially decreases when the tunnel barrier height increases, in agreement with Ref.~\cite{btk} for an s-wave SC. These results remain valid as long as $L_S>>\xi$, with $\xi$ the localization length of the superconductor~\cite{Klinovaja2012}, when LAR is the dominant process. Interestingly, we observe a crossing point where $A^L_{\downarrow \downarrow}$ drops below $A^L_{\downarrow \uparrow}= A^L_{\uparrow \downarrow}$ [see inset of Fig~\ref{fig2}(b)]. The latter reaches a plateau for larger $U_R$ values, indicating a residual contribution from usual, as opposed to the (quasi-) zero-energy-mediated LAR. In Figs.~\ref{fig1}(b),(c), and~\ref{fig2}(c),(d), we normalize the conductances by their respective values at $U_R=U_L$. The values of $G_{\alpha \alpha}(U_R=U_L)$ are shown in table I of Appendix \ref{S1} and are within the reach of standard state-of-the-art experimental techniques~\cite{wang2022parametric,dvir2023realization}.

\subsection{$G_{LL}$ and $G_{RR}$ as probes for topological phases}

The dependence of $G_{LL}$ and 
$G_{RR}$ on $U_R$ can be used to distinguish between trivial and topological phases. In Figs.~\ref{fig1}(b) 
and (c), respectively, we compare $G_{LL}$ and $G_{RR}$ as functions of $U_R/U_L$ for the trivial (dashed 
lines) and topological (solid lines) regimes for the pristine wire, in the presence of a smooth 
nonuniform $\mu_S(x)$ (quasi Majorana), 
and moderate disorder (with $W = 3.5 \Delta$ and averaging over $100$ disorder realizations)~\cite{disorderDasSarma}. 
Notice that \textit{only in the topological phase} $G_{RR}$ exhibits a similar dependence on $U_R/U_L$ 
as $G_{LL}$. Strikingly, this result remains true in the presence of inhomogeneities in the wire (green and orange 
solid lines), with $G_{LL}\sim G_{RR}$ qualitatively reproducing the nonlocal behavior of the clean 
case, which signals the robustness of the topological phase. In the trivial phase, quasi Majoranas and 
disorder-driven states do not reproduce this feature; $G_{LL}$ and $G_{RR}$ are uncorrelated 
in this case (dashed lines). These results also hold for symmetric and asymmetric $\mu_S(x)$ profiles. 
The nonlocal suppression of LAR leading to $G_{LL}\sim G_{RR}$ follows from the strong 
correlation (hybridization $\varepsilon_m$) between MZM wave functions, absent in trivial ABS as discussed in 
Appendix \ref{validityApp} (see Fig.~\ref{EmergenceExtendedABS} and text after Fig.~\ref{LeftvsRight symmetry}). This leads to a symmetric response of the local conductances in the topological case to variations of the gate potentials about $U_L=U_R$ and, ultimately, to $G_{LL}\sim G_{RR}$. As shown in Figs.~\ref{fig1}(b) 
and (c) (symbols) [see also Fig.~\ref{fig5}(a)], this unique correlation of the local conductances holds for both $T=0$ and electron temperatures $T<20$ mK.

\subsection{Identifying fine-tuned trivial peaks}

The nonlocality of LAR is an experimentally accessible test of whether left and right zero-bias peaks are correlated. A mere coincidental appearance of zero-bias peaks at similar regions in parameter space is insufficient as a signature of MZMs. To illustrate this point, we show in Figs.~\ref{fig2}(c) and (d) that fine-tuned \textit{nontopological} ABSs (pink dotted curves) yield $G_{LL}\neq G_{RR}$ as functions of $U_R/U_L$, consistent with the results for the trivial phase. These nontopological ABSs are symmetrically localized in the normal parts of the wire, Fig. \ref{fig1}(a), which act as quantum wells,  and perfectly mimic the zero-bias peak characteristic of MZMs, i.e., $G_{LL}=G_{RR}= 1.99 e^2/h$ at $U_R=U_L$~\footnote{The nontopological ABSs here are engineered following Ref.~\cite{NonLocalLoss}. We set the spin-orbit coupling $\alpha_R$ to zero within the superconducting region of the nanowire and choose $V_z = 1.16 V_c$ and $\mu_N = 2.58 meV$.}. Upon varying $U_R/U_L$, one can tell apart these accidental ABSs ($G_{LL}\neq G_{RR}$) from true topological states ($G_{LL}\sim G_{RR}$). 
We point out that for significantly smaller wires ($L_S < \xi$), such that the trivial ABSs hybridize, the pink curves could, in principle, mimic the topological case. However, since these ABSs are fine-tuned, the conductance is not robust against variations of the parameters, e.g., $\mu_N$ and $V_z$. For wires of length $2 ~\rm{\mu m}$ or longer, currently used in experiments~\cite{nayakExperiment,frolovscipost}, our proposal for distinguishing topological from trivial phases by measuring only $G_{LL}$ \textit{and} $G_{RR}$ could be implemented. 
\begin{figure}[t]
	\centering
	\includegraphics[width=0.238\textwidth]{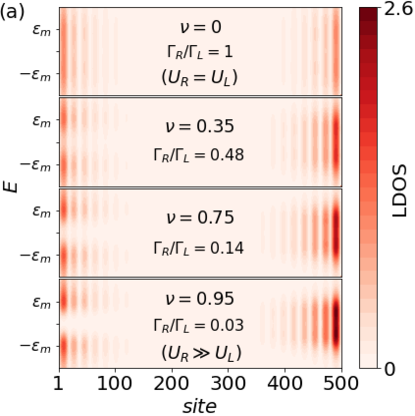}
	\hfill
	\centering
	\includegraphics[width=0.238\textwidth]{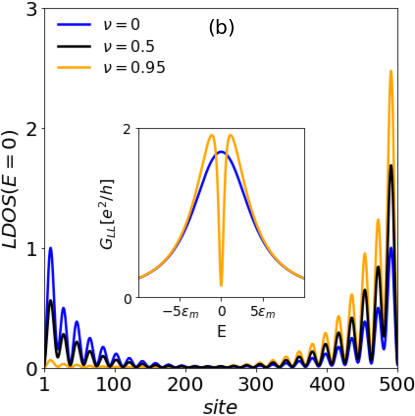}
	\caption{(a) LDOS (color map) as a function of $E$ for different values of $\nu$. Around zero energy, a dip emerges in the LDOS on the left side. On the right side, the LDOS is enhanced  at zero energy as $\nu$ increases ($\Gamma_R \rightarrow 0$). (b) LDOS at $E = 0$ for different values of $\nu$. The local conductance (inset) follows the dip in the LDOS. The nanowire parameters are the same as those used in Fig.~\ref{fig2}, $V_z=1.40V_c$,  and $\Gamma_L = 1.6 \Delta$. We consider $\varepsilon_m$ as the energy of the lowest mode. The values of the LDOS were normalized by their maximum at $\nu = 0$.}
	\label{fig3}
\end{figure}

\section{Local density of states} \label{LDOS section}

The suppression of the (quasi-) zero-energy mediated LAR observed in the topological phase of Fig.~\ref{fig2}(b) can be explained by the emergence of a zero-energy dip in the LDOS at the left side of the wire as $\Gamma_R$ decreases (i.e., $U_R$ increases), Fig.~\ref{fig3}(a). We obtain the LDOS from the imaginary part of the retarded Green function~\cite{dattaBook} within the Bogoliubov-de Gennes formalism~\footnote{In agreement with the LDOS obtained using the package KWANT~\cite{kwant}.}, $\mathcal{G}^r = (E - H_{BdG} - \Sigma)^{-1}$, where $\Sigma$ is the self-energy.

In Fig.~\ref{fig3}(a) we show the LDOS (color map) as a function of the energy $E$ for $V_z=1.40 V_c$. We observe a suppression (enhancement) of the LDOS around $E = 0$ on the left (right) side of the wire for large values of the lead-asymmetry $\nu=(\Gamma_L - \Gamma_R)/(\Gamma_L + \Gamma_R)$.  
More specifically, we show in Fig.~\ref{fig3}(b) the LDOS at $E = 0$ for several values of $\nu$. At $\nu=0$ ($U_R=U_L \Leftrightarrow \Gamma_R=\Gamma_L$, blue curve), the LDOS has equal weights on the left and right ends of the wire. As $\Gamma_R$ decreases (i.e., $U_R$ increases) for a fixed $\Gamma_L$ ($U_L$), black and orange curves, an asymmetry develops between the left and right sides with the LDOS building up on the right side. The local conductance $G_{LL}$, which we calculate here analytically for the effective model (see Appendix \ref{lowenergyH}), shows a suppression as $\Gamma_R$ is decreased [see inset in \ref{fig3}(b)]. This suppression is similar to those in Figs.~\ref{fig1}(b), (c) and \ref{fig2}(c), (d) as $U_R/U_L$ is increased. In the topological phase, the local conductances are correlated, i.e., $G_{LL}\sim G_{RR}$, but suppressed for different reasons. While $G_{LL}$ is reduced due to the emergence of a dip in the LDOS on the left end of the wire ($U_L$ fixed), $G_{RR}$ is suppressed because the right tunnel barrier $U_R$ increases, even though the LDOS on the right end of the wire is correspondingly enhanced. Thus, the lead-asymmetry $\nu$, controlled by $U_L$ and $U_R$, allows us to manipulate the LDOS. Our analysis agrees with the result of Ref.~\cite{flensbergMajoranaChain}, where it is shown that for one lead, i.e., $\Gamma_R=0 \Leftrightarrow \nu = 1$, the conductance vanishes for an even number of MZMs.

A complementary analysis using non-Hermitian topology~\cite{aguadoNHT} is provided in Appendix \ref{nonHermitianSection}. Here, the asymmetry in the LDOS can be understood by a bifurcation of the imaginary part of the energy of the lowest mode and its particle-hole partner. This means that one of the Majorana components leaks to the lead, causing the zero-energy dip in the LDOS, while the other has its lifetime increased. It is worth mentioning that after the bifurcation (exceptional point), the real part of the energy of the lowest mode is exactly zero, suggesting that it is possible to artificially create non-hybridized Majorana-like zero modes in finite wires by exploring the lead-asymmetry~\cite{aguadoNHT}. However, the suppression of $G_{RR}$ due to the barrier height shows that this is not a true zero-energy Majorana mode, which instead would present a universal zero-bias peak of $2 e^2/h$~\cite{PLeeMajoranaInducedAndreev}. In passing, we mention that, in principle, our results can also be analyzed within the scattering theory of topological invariants~\cite{akhmerov2011quantized}.

\section{Majorana-based transistor}

\begin{figure}
	\centering
	\includegraphics[width=0.35\textwidth]{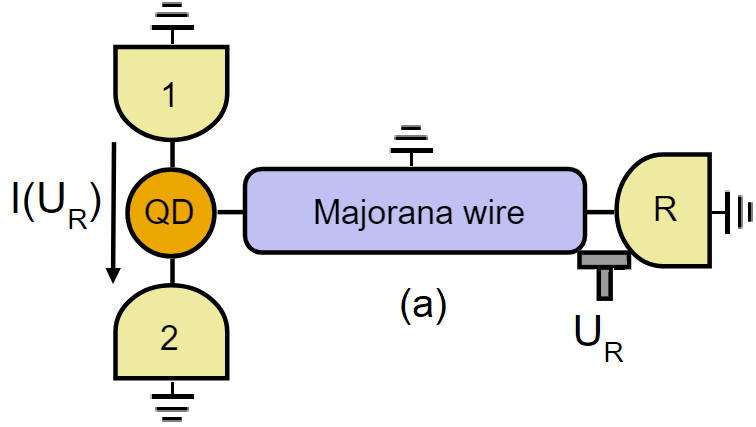}
	\quad
	\centering
	\includegraphics[width=0.235\textwidth]{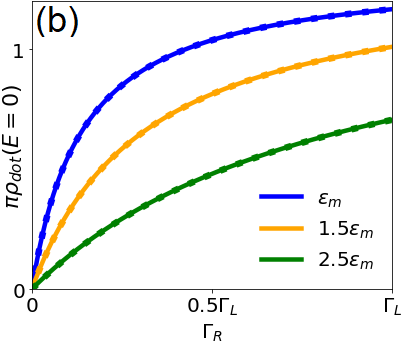}
	\hfill
	\centering
	\includegraphics[width=0.235\textwidth]{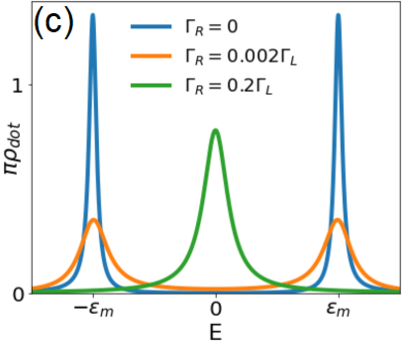}
	\caption{(a) Setup of the Majorana-transistor: QD coupled to leads $1$ and $2$ and to a Majorana wire connected to a third lead $R$. (b) DOS of the QD for different values of $\Gamma_R$, solid lines. At zero energy, $\rho_{dot} \propto F$, the latter is shown as dotted lines. The control over the coupling to the right lead ($U_R$) allows us to manipulate the QD DOS, thus suppressing or enhancing it around zero energy. (c) DOS of the QD at $E = 0$ as a function of $\Gamma_R$ for different $\varepsilon_m$. At $\Gamma_R = 0$, the zero-bias peak vanishes.  The parameters used are $t_0 = \Gamma_L = 100 \varepsilon_m$, and $\epsilon_d = 53.3 t_0$.}
	\label{FigTransistor}
\end{figure}

The manipulation of the LDOS via lead asymmetry allows us to gate control the currents in multi-terminal devices. To illustrate this, we propose a Majorana-based transistor using a QD coupled to leads 1 and 2, and to a Majorana wire~\cite{barangerDot, vernekLeakage}, Fig.~\ref{FigTransistor}(a). The right side of the topological wire is then coupled to a third lead that acts as a ``base gate'' ($U_R$), controlling the current flow through the QD. For simplicity, we consider a spinless effective model since the MZMs only couple to one spin direction~\cite{PLeeSESAR}. Experimentally, this polarization can be realized via a global Zeeman field. 

The Hamiltonian of the system is given by~\cite{demler, PLeeMajoranaInducedAndreev, barangerDot, flensbergDot}

\begin{equation} \label{Hamiltonian Transistor}
    H = i \varepsilon_m \gamma_1 \gamma_2 + \epsilon_d c^\dagger_d c_d + t_0 \left(c_d^\dagger - c_d \right) \gamma_1 + H_T,
\end{equation}
with $\gamma_i = \gamma_i^\dagger$ the Majorana operators, $\varepsilon_m$ the hybridization energy, $\epsilon_d$ the dot level (tunable via an external gate), and $c_d$ ($c_d^\dagger$) the annihilation (creation) operator of an electron in the QD. The hopping parameter $t_0$ couples the QD to $\gamma_1$ and $H_T$ corresponds to the tunneling Hamiltonian between the system and the external leads,
\begin{equation}
    H_T =  \sum_{k, i = 1, 2} t_{k,i} ( c^\dagger_d d_{k, i} + H.c.) + \sum_k t_{k, R} (d_{k, R}^\dagger - d_{k, R} ) \gamma_2,
\end{equation}
where $t_{k, j}$ is the hopping amplitude and $d_{k, j}$ ($d_{k, j}^\dagger$) annihilates (creates) an electron with momentum $k$ on lead $j$ . Note here that $t_{k, R} = t_{k, R} (U_R)$.

To obtain the LDOS, we first calculate the retarded Green function (GF) of the system using the basis $\{c_d, c_d^\dagger, \gamma_1, \gamma_2\}$,
\begin{equation}
	\mathcal{G}^r(E) = \begin{pmatrix}
 	\mathcal{G}^r_{\rm{dot}} & \Tilde{F} \\
 	\Tilde{F} & \mathcal{G}^r_{\rm{MW}} \end{pmatrix}
\end{equation}
where $\mathcal{G}^r_{\rm{dot}}$ and $\mathcal{G}^r_{\rm{MW}}$ are the GFs of the QD and Majorana wire, respectively. Since the leaking of the Majorana mode into the QD can be characterized by $ \rho_{\rm{dot}} = - \frac{1}{\pi} \text{Im}[\mathcal{G}^r_{\rm{dot}} ]$, here we mostly focus on $\mathcal{G}^r_{\rm{dot}}$, written as
\begin{equation}
    \mathcal{G}^r_{\rm{dot}} = \begin{pmatrix}\mathcal{G}_e & F \\ F & \mathcal{G}_h \end{pmatrix},
\end{equation}
with $\mathcal{G}_e$ and $\mathcal{G}_h$ the electron and hole components, respectively, and $F$ the so-called anomalous part. Up to first order in the energy $E$~\footnote{Here we are interested in the low-energy regime, $E \sim  0$.}, we obtain
\begin{equation} \label{Geh transistor}
	\mathcal{G}_{e(h)} = \tilde{Z}^{-1}\left[-\varepsilon_m^2 (\pm \epsilon_d + i \Gamma_L + E) + E ( \pm  i \epsilon_d \Gamma_R  - \Gamma_L \Gamma_R) \right] + F, 
\end{equation}
\begin{equation} \label{F transistor}
	F =  -t_0^2 (E + i \Gamma_R) \tilde{Z}^{-1},
\end{equation}
where $+ (-)~\epsilon_d$ refers to electrons (holes), $\Gamma_L=(\Gamma_1 + \Gamma_2)/2$ is the effective coupling between the QD and leads $1$ and $2$, with $\Gamma_i=2\pi|t_i|^2\rho_i$ and $\rho_i$ the DOS of lead $i$, and
\begin{eqnarray} \label{Z transistor}		
	\tilde{Z} &=& \varepsilon_m^2 (\epsilon_d^2 + \Gamma_L^2 - 2i \Gamma_L E) + 2 \Gamma_L \Gamma_R t_0^2  \nonumber \\
	&-&i E \left[ \Gamma_R (\epsilon_d^2 + 
			\Gamma_L^2) + 2 t_0^2 (\Gamma_L + \Gamma_R) \right].
\end{eqnarray}
A detailed derivation of the equations above can be found in Appendix~\ref{AppTransistor}. We now analyze some limiting cases. 

For $\varepsilon_m = 0$, $\rho_{\rm{dot}}(E=0) = 1/2\pi \Gamma_L$, regardless of $\Gamma_R$ and $\epsilon_d$, in agreement with our numerical results, which show that as $\varepsilon_m \to 0$, the left side of the wire becomes insensitive to variations in $\Gamma_R$, see, for instance, the cyan curve in Fig. \ref{fig2}(b). In finite wires, however, there is, in general, a finite hybridization energy. Hence, in what follows we assume $\varepsilon_m \neq 0$, $\epsilon_d \gg \Gamma_L \sim t_0 \gg \varepsilon_m$, and take $E \to 0$.

If the right lead is pinched off, the conductance is not mediated by the MZM. This can be seen from Eqs.~\eqref{Geh transistor}-\eqref{Z transistor}, in which, for $\Gamma_R = 0$, $\mathcal{G}_{e(h)} \approx (\mp \epsilon_d -i\Gamma_L)/\epsilon_d^2 $. Hence, the entire contribution to the LDOS comes from the distant dot level. This result agrees with the LDOS discussion from the previous section \ref{LDOS section}. Since $\gamma_2$ is uncoupled from the right lead, this characterizes the situation with $\nu = 1$. This highly asymmetric configuration enforces a zero-energy dip in the LDOS of the Majorana wire, as shown in Fig.~\ref{fig3}(a) (last panel). Therefore, the conductance is negligible, with no contribution from the Majorana mode.

The right lead unlocks the leakage of the Majorana mode into the QD. We start by weakly coupling  the wire to the right lead, $\Gamma_R \ll \varepsilon_m$. In this case, we obtain $F \approx \Gamma_R t_0^2/\varepsilon_m^2 \epsilon_d^2$ and
\begin{equation}\label{rhoWeakGR}
    \rho_{\rm{dot}} \approx \frac{\Gamma_L}{\pi\epsilon_d^2} + \frac{\Gamma_R t_0^2}{\pi\varepsilon_m^2\epsilon_d^2},
\end{equation}
where we identify the first term as the residual contribution of the dot level and the second as the Majorana contribution enabled by $\Gamma_R$. If we now take $\Gamma_R >> \varepsilon_m^2\epsilon_d^2/t_0^2$ (strong coupling), 
\begin{equation} \label{rhoLargeGR}
    \mathcal{G}_{e(h)} \approx F \approx -i/2\Gamma_L, \,\, \rho_{\rm{dot}} = 1/2\pi \Gamma_L,
\end{equation}
which is precisely the value of the DOS of the dot for the infinite wire ($\varepsilon_m = 0$). Interestingly, as long as $\varepsilon_m$ is the smallest energy scale, $\rho_{\rm{dot}} \propto -\text{Im}[F]$. This is shown in Fig. \ref{FigTransistor}(b); the dotted (Im[$F$]) and solid ($\rho_{\rm dot}$) lines fall on top of each other. The Majorana wire induces superconductivity on the dot so that the Majorana leakage comes entirely from the anomalous term of the GF. Figure~\ref{FigTransistor}(b) also shows that the larger the hybridization energy, the larger the coupling to the right lead has to be to converge to the value of $\rho_{\rm{dot}} (\Gamma_R \gg \epsilon_d^2)$, which depends only on $\Gamma_L$, in agreement with Eqs.~\eqref{rhoWeakGR} and \eqref{rhoLargeGR}. In Fig. \ref{FigTransistor}(c), we show $\rho_{\rm{dot}} $ as a function of $E$. We observe that only when $\Gamma_R$ is increased, a peak around $E = 0$ emerges. 

Finally, using a symmetrical potential drop between leads $1$ and $2$, i.e. $V = V_L/2 = - V_R/2$, and defining the conductance through the dot as $G = dI_1/dV$, we obtain
\[
G = \begin{cases}
			0, & \text{for } \Gamma_R = 0, \\
            e^2/2h, & \text{for } \Gamma_R >> \varepsilon_m^2\epsilon_d^2/t_0^2.
		 \end{cases}
\]
This result highlights the role of the tunnel voltage $U_R$ ($\Gamma_R$) in the control of the current, which characterizes a transistor.  

\section{Temperature Effects}

\begin{figure}[htb!]
	\centering    \includegraphics[width=0.225\textwidth]{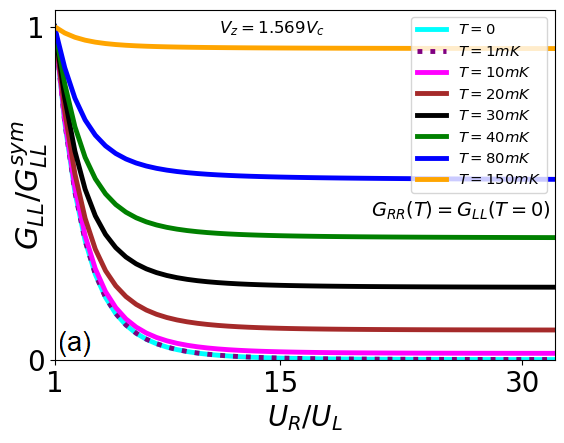}
	\hfill
	\centering
	\includegraphics[width=0.25\textwidth]{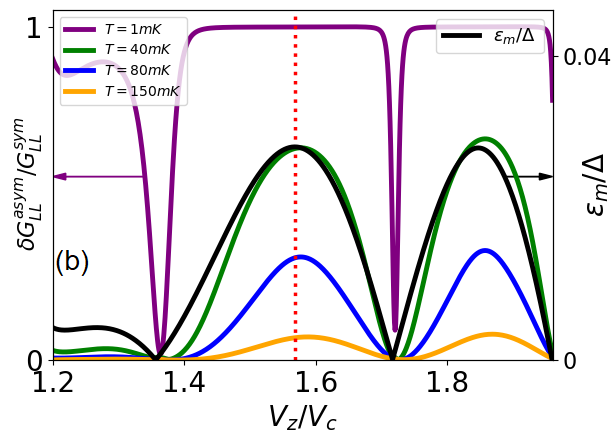}
	\caption{Temperature effects on $G_{LL}$, $G_{RR}$, and $\delta G^{asym}_{LL}$ in the topological phase. (a) Local conductances $G_{LL}$ and $G_{RR}$ as functions 
			of $U_R/U_L$ for $T = 0, 1, 10, 30, 20, 40, 80, 150$ mK and $V_z/V_c=1.569$ for a disordered wire (single realization). 
			(b) Hybridization energy $\varepsilon_m$
			and asymmetric conductance deviation $\delta G^{asym}_{LL}$ as functions of 
			$V_z/V_c>1$. The dotted red line in (b) corresponds to the $V_z/V_c$ in (a) that maximizes 
			$\varepsilon_m$. For $T < 20$ mK $G_{LL} \sim G_{RR}$, first decreasing then 
			attaining a plateau as $U_R/U_L$ increases, similarly to Figs.~\ref{fig1}(b), (c). The magnitude of the plateaus 
			in $G_{LL}$ strongly depends on $\varepsilon_m$ and $T$. In contrast, $G_{RR}$ vs. $U_R/U_L$ is insensitive to
			$T$, maintaining its $T=0$ behavior. For $T=30$, 40 meV, $G_{LL}$ and $G_{RR}$  become less correlated (red and green lines).
			For even higher temperatures  $ 40<T<150$ mK, $G_{LL}$ still displays nonlocal effects as 
			quantified by the asymmetric deviation $\delta G^{asym}_{LL}$, which shows sizable oscillations as a function of $V_z/V_c$. These 
			modulations in  $\delta G^{asym}_{LL}$ are commensurate with the Majorana oscillations in 
			$\varepsilon_m$ as shown in (b) (black line); they provide an additional  
			probe for the topological phase. Parameters: same as in Fig.~\ref{fig1}. }
	\label{fig5}
\end{figure}

We now analyze how temperature effects modify our previous results.  Finite temperatures tend to suppress the correlation between $G_{LL}$ and $G_{RR}$, as shown in 
Fig.~\ref{fig5}(a). This occurs because $G_{LL}$ and $G_{RR}$ are affected 
distinctively by the interplay of the LDOS and the Fermi function $f(\varepsilon)$. More specifically,  
in a linear-response conductance calculation, $\partial f(\varepsilon)/ \partial \varepsilon$ 
(width $\sim k_BT$) multiplies the LDOS, i.e., 
$G_{LL} \sim \partial f(\varepsilon)/\partial \varepsilon \rho_L(\varepsilon)$ and  
$G_{RR} \sim \partial f(\varepsilon)/\partial \varepsilon \rho_R(\varepsilon)$, where $\rho_L$ ($\rho_R$) 
denotes the left (right) LDOS. As shown in Fig.~\ref{fig3}(b), while 
$\rho_L(\varepsilon)$ develops a dip of width $\varepsilon_m$ near zero energy as the left-right asymmetry parameter $\nu$ 
increases, $\rho_R(\varepsilon)$ is in contrast slightly enhanced and essentially uniform in $\varepsilon$. For increasing temperatures, the Lorentzian-shaped 
$\partial f(\varepsilon)/ \partial \varepsilon$ samples wider energy ranges in  
both $\rho_L$ and $\rho_R$ near zero energy.  As it turns out, $G_{LL}$ vs. $U_R/U_L$ is 
strongly dependent on temperature variations, see curves for $T=40, 80, 150$ mK in Fig~\ref{fig5}(a), while $G_{RR}$ is mostly insensitive 
(here all curves lie on top of the $T=0$ purple line). The competition between 
$\varepsilon_m$ and $k_BT$ determines the behavior of $G_{LL}$ vs. $U_R/U_L$ and, in particular, the magnitude of the 
attained plateau for $U_R\gg U_L$, see Appendix \ref{effectiveModelAndTemp} for a detailed analysis.  When $\varepsilon_m \gg k_BT$, which holds in Figs.~\ref{fig1}(b), \ref{fig1}(c), and \ref{fig5}(a) for $T< 20 $ mK and nanowires of $\sim 2~\mu$m, $G_{LL} \sim G_{RR}$. 
Recent experiments report $T \sim 30-40$ mK for measured electron temperatures~\cite{heedt2021shadow,nayakExperiment,dvir2023realization, van2023electrostatic}.
For higher $T$'s the $\varepsilon_m \gg k_BT$ no longer holds and $G_{LL}$ and $G_{RR}$  become less and less correlated, 
cf. curves for $T=30, 40, 80, 150$ mK in Fig.~\ref{fig5}(a). However, as we discuss next, the asymmetric deviation 
$\delta G^{asym}_{LL} = G_{LL}^{U_R = U_L} - G_{LL}^{U_R \gg U_L}$ vs. $V_z/V_c$ can still be used to identify the topological phase at these higher temperatures (see also Appendix~\ref{GLLoscillationsWithVz}).

In Fig.~\ref{fig5}(b) we show that for finite temperatures, 
$\delta G^{asym}_{LL}$
displays modulations as a function of $V_z/V_c$ that are commensurate with Majorana oscillations, 
quantified by $\varepsilon_m$ (solid black curve). This feature is more 
clearly visible as oscillations matching $\varepsilon_m$ in the higher temperature range  
$30<T<150$ mK. Hence, the modulations in $\delta G^{asym}_{LL}$ provide a 
direct measure of Majorana oscillations as probed by the nonlocality of $G_{LL}$ at 
surprisingly high $T$'s. In addition, the maxima of $\delta G^{asym}_{LL}$ can be used 
to find the optimal $V_z/V_c$ for which $\varepsilon_m$ is the largest. This ratio can then be used at low $T$ to enhance the correlation 
between $G_{LL}$ and $G_{RR}$ [this was done in Figs.~\ref{fig1}(b) and (c) for the pristine case, and Fig. \ref{fig5}(a)]. Interestingly, even though trivial ($V_z/V_c<1$) quasi-majorana or fine-tuned ABS modes can eventually mimic Majorana oscillations in $G_{LL}$, the asymmetric deviation $\delta G^{asym}_{LL}$, on the other hand, does not display modulations in these cases.

As discussed, to observe the nonlocal suppression of the local conductances the ideal regime is $\varepsilon_m \gg k_B T$. This limits the length of the wire for our proposed Majorana transistor. This particular application, however, does not rely on a specific realization for the central Majorana wire, which could, for instance, be implemented via an artificial Kitaev chain~\cite{dvir2023realization, tsintzis2022creating, bordin2024signatures} such as a $3$-quantum dot array, with an additional QD as a probe, similarly to the setup studied in Ref. \cite{souto2023probing}. In this case, the control over the hybridization $\varepsilon_m$ can be realized via external gates and the regime in which temperature effects are negligible can be enabled.

\section{Protocol using only the local conductance}

We propose a simplified protocol consisting of only local conductance measurements. 
(i) One should vary the tunnel-barrier ratio $U_R/U_L$ ($U_L$ kept fixed) and check whether the local conductances exhibit approximately 
equal suppressions, i.e., $G_{LL} \sim G_{RR}$, as $U_R/U_L$ increases, similar to Fig.~\ref{fig1}(b),(c). This response should be the same (i.e., symmetric) when one increases $U_L/U_R$ ($U_R$ kept fixed). (ii) One should measure the asymmetric deviation $\delta G^{asym}_{LL}$ as 
a function of the Zeeman field to see whether it is modulated with well-defined zeros, similar to Fig.~\ref{fig5}(b), thus signaling Majorana 
oscillations.  If either (i) or (ii) or both hold true and are robust upon variations of the system parameters~\cite{nayakExperiment}, the 
protocol has identified a topological phase. The above procedure -- with its unique nonlocal features -- could also be used as a novel diagnostic tool in combination with the so-called Topological Gap Protocol~\cite{nayakExperiment} to identify a topological phase. 

\section{Validity of the proposed protocol}

In this section, we elaborate on the validity of our proposed protocol by investigating several types of effects that can be detrimental to our proposal (Appendices~\ref{validityApp} and \ref{FiniteBiasSec}).

Inhomogeneities in the parameters of Eq. \eqref{Hcontinuum}, in particular, charge impurities \cite{disorderDasSarma}, are one of the main obstacles to the detection of MZMs. We have taken this effect into account by including disorder in the chemical potential in Figs. \ref{fig1} and \ref{fig5}. For these results, we set $W = 3.5 \Delta$, which is a moderate disorder strength \cite{disorderDasSarma}. In Appendix \ref{validityApp}, we consider larger values of disorder and analyze when our proposal breaks down. Our results indicate that our protocol remains viable for $W \lesssim 10.5 \Delta$. At this point, the correlation between the local conductances is partially lost even at $T=0$, see Fig.~\ref{disorder}, similarly to the quasi-Majorana case shown in Figs.~\ref{fig1}.

We note that, even in the moderate disorder case, there may be rare disorder realizations having correlated zero-bias peaks (due to hybridization of the trivial modes); these are irrelevant if self-averaging holds, otherwise, these zero-energy modes fall under ``fine-tuned'' cases because they move away from the zero-energy when some other parameter is varied. We also analyze the emergence of extended ABSs as both disorder and wire lengths are varied, Fig.~\ref{EmergenceExtendedABS}. We conclude that long wires are less susceptible to the emergence of trivial extended states, see Appendix \ref{validityApp} for details. Finally, we mention that in Ref. \cite{laubscher2023majorana}, germanium hole nanowires are considered potential platforms to reduce disorder in Majorana wires, which could be used to avoid trivial extended ABSs.

We have also introduced other possible detrimental effects in our setup, such as (i) nontopological superconducting nanowire segments of length $\ell$, (ii) normal ($\Delta=0$) nanowire sections of length $\ell_N$, and (iii) potential barriers of different lengths, see Appendix \ref{validityApp} for details. We have verified that our protocol is robust against those types of defects and remains valid provided that $\ell, \ell_N \lesssim  0.6 L_S $. We acknowledge, however, that this is not an exhaustive list of undesirable effects.

\section{Conclusions}

We have identified nonlocal effects on the zero-bias local conductances 
$G_{LL}$ and $G_{RR}$ vs. the tunnel-barrier ratio $U_R/U_L$ as unique probes, within a protocol, for 
topological phases in three-terminal hybrid nanowires. 
For $T< 20$ mK, $G_{LL} \sim G_{RR}$ only in the topological phase, while 
for $T=30$, $40$ mK, their correlation becomes weaker. 
At higher temperatures $40<T<150$~mK, we predict that the asymmetric deviation 
$\delta G^{asym}_{LL}= G_{LL}^{U_R = U_L} - G_{LL}^{U_R \gg U_L}$ vs. $V_z$ 
shows modulations commensurate with Majorana oscillations as an additional signature of 
the topological phase. 

To understand the mechanism behind the local conductance ($G_{LL}$) suppression upon disconnecting a distant metallic lead ($\Gamma_R$), we have investigated the LDOS and found that a dip in the zero-energy local density of states, $\rho (E = 0, i = 0)$, causes the conductance drop ($G_{LL} \to 0$). We have then exploited this nonlocal control over the LDOS and proposed a Majorana-based transistor, where the current flowing through a side-coupled quantum dot is controlled by tuning the coupling to the right lead.

Finally, we have analyzed several detrimental effects such as temperature, disorder, and other inhomogeneities in the Majorana nanowire in order to provide realistic parameter regimes where the nonlocalities of local conductances due to the presence of MZMs can be tested.

\begin{acknowledgments} We acknowledge discussions with D. Loss, J. Klinovaja, H. F. Legg, and R. Hess. R.A.D acknowledges discussions with G. P. Mazur and L. Pupim. This work was supported by the São Paulo Research Foundation (FAPESP), Grant No. 2020/00841-9, and Conselho Nacional de Pesquisas (CNPq), Grant No. 306122/2018-9. We also acknowledge support from the Swiss NSF.
\end{acknowledgments}

\appendix

\section*{Appendix}

\section{Semiconducting nanowire: conductance simulations}
\label{S1}

For the numerical calculation, we discretize the continuum BdG Hamiltonian in Eq.~(1) of the main text and obtain the tight binding model~\cite{disorderDasSarma}

\begin{equation} \label{HTB}
\begin{split}
H_{\mathrm{TB}}=& \sum_{j=1}^{N-1}\left[-t|j+1\rangle\left\langle j\left|\tau_{z}+i \alpha \right| j+1\right\rangle\langle j| \sigma_{y} \tau_{z}+\text { H.c. }\right] 
+ \\ 
&+\sum_{j=1}^{N}\left[\Delta|j\rangle\left\langle j\left|\tau_{x}+\left(2 t-\mu_{j}\right)\right| j\right\rangle\left\langle j\left|\tau_{z}+V_{Z}\right| j\right\rangle\langle j| \sigma_{x}\right],
\end{split}
\end{equation}
where $t = \hbar^2/2 m^* a^2$ and $\alpha = \alpha_R/2a$, with $a$ being the ``lattice'' parameter. In the main text, we use $a = 5~\rm{nm}$~\cite{NonLocalLoss}.

For convenience, the conductances $G_{LL}$ and $G_{RR}$ in Figs.\ref{fig1} (b), (c), and \ref{fig2} (c), (d) were normalized by their respective values at $U_R = U_L$, $G^{\rm{sym}}_{LL}$ and $G^{\rm{sym}}_{RR}$. In Table~\ref{Table1}, we show the actual values of $G^{\rm{sym}}_{LL}$ and $G^{\rm{sym}}_{RR}$ for the pristine $2.5 \mu m$ wire and in the presence of a smooth confining potential at $V_z/V_c = 0.10, 0.95, 1.15$, and $1.40$. For the pristine wire, we also show the case $V_z/V_c = 1.20 $ and the fine-tuned ABS in a nontopological nanowire (no spin-orbit coupling in the superconducting part of the wire)~\cite{NonLocalLoss}.

\begin{table}[htb]
\centering
\begin{tabular}{|p{2.7cm}|p{2.7cm}|p{2.7cm}| }
\hline
\multicolumn{3}{|c|}{Pristine wire} \\
\hline
 $\,\,\,\,\,\,\,\,\,\,\,\,\,\,\,\,\,\,\,\, V_z/V_c$ & $G_{LL} (U_R = U_L)[e^2/h]$ & $G_{RR} (U_R = U_L) [e^2/h]$ \\
\hline
 \,\,\,\,\,\,\,\,\,\,\,\,\,\,\,\,\,\,\,\, 0.10  & $\,\,\,\,\,\,\,\,\,\,\,\,\,\,\,\,\,\,\,\,\,\,0.012$ & $\,\,\,\,\,\,\,\,\,\,\,\,\,\,\,\,\,\,\,\,\,\,0.012$ \\
 \,\,\,\,\,\,\,\,\,\,\,\,\,\,\,\,\,\,\,\, 0.95  & $\,\,\,\,\,\,\,\,\,\,\,\,\,\,\,\,\,\,\,\,\,\,0.038$ & $\,\,\,\,\,\,\,\,\,\,\,\,\,\,\,\,\,\,\,\,\,\,0.038$ \\
 \,\,\,\,\,\,\,\,\,\,\,\,\,\,\,\,\,\,\,\, 1.15  & $\,\,\,\,\,\,\,\,\,\,\,\,\,\,\,\,\,\,\,\,\,\,1.997$ & $\,\,\,\,\,\,\,\,\,\,\,\,\,\,\,\,\,\,\,\,\,\,1.997$ \\
 \,\,\,\,\,\,\,\,\,\,\,\,\,\,\,\,\,\,\,\, 1.20  & $\,\,\,\,\,\,\,\,\,\,\,\,\,\,\,\,\,\,\,\,\,\,1.999$ & $\,\,\,\,\,\,\,\,\,\,\,\,\,\,\,\,\,\,\,\,\,\,1.999$ \\
 \,\,\,\,\,\,\,\,\,\,\,\,\,\,\,\,\,\,\,\, 1.40  & $\,\,\,\,\,\,\,\,\,\,\,\,\,\,\,\,\,\,\,\,\,\,1.995$ & $\,\,\,\,\,\,\,\,\,\,\,\,\,\,\,\,\,\,\,\,\,\,1.995$ \\
 \,\,\,\,\,\,\,\,\,\,\,\, NT -- ABS & $\,\,\,\,\,\,\,\,\,\,\,\,\,\,\,\,\,\,\,\,\,\,1.991$ & $\,\,\,\,\,\,\,\,\,\,\,\,\,\,\,\,\,\,\,\,\,\,1.991$ \\
\hline

\multicolumn{3}{|c|}{Smooth confining potential} \\
\hline
 $\,\,\,\,\,\,\,\,\,\,\,\,\,\,\,\,\,\,\,\, V_z/V_c$ & $G_{LL} (U_R = U_L)[e^2/h]$ & $G_{RR} (U_R = U_L)[e^2/h]$ \\
\hline
 \,\,\,\,\,\,\,\,\,\,\,\,\,\,\,\,\,\,\,\, 0.10  & $\,\,\,\,\,\,\,\,\,\,\,\,\,\,\,\,\,\,\,\,\,\,0.015$ & $\,\,\,\,\,\,\,\,\,\,\,\,\,\,\,\,\,\,\,\,\,\,0.012$ \\
 \,\,\,\,\,\,\,\,\,\,\,\,\,\,\,\,\,\,\,\, 0.95  & $\,\,\,\,\,\,\,\,\,\,\,\,\,\,\,\,\,\,\,\,\,\,0.178$ & $\,\,\,\,\,\,\,\,\,\,\,\,\,\,\,\,\,\,\,\,\,\,0.044$ \\
 \,\,\,\,\,\,\,\,\,\,\,\,\,\,\,\,\,\,\,\, 1.15  & $\,\,\,\,\,\,\,\,\,\,\,\,\,\,\,\,\,\,\,\,\,\,1.997$ & $\,\,\,\,\,\,\,\,\,\,\,\,\,\,\,\,\,\,\,\,\,\,1.997$ \\
 \,\,\,\,\,\,\,\,\,\,\,\,\,\,\,\,\,\,\,\, 1.40  & $\,\,\,\,\,\,\,\,\,\,\,\,\,\,\,\,\,\,\,\,\,\,1.927$ & $\,\,\,\,\,\,\,\,\,\,\,\,\,\,\,\,\,\,\,\,\,\,1.927$ \\
\hline

\end{tabular}
\caption{Local conductances $G_{LL}$ and $G_{RR}$ at $U_L=U_R$ for the pristine wire and in the presence of a smooth confining potential. The parameters are detailed in Fig. 1 of the main text.}
\label{Table1}
\end{table}

\begin{figure}[htb]
	\centering
	\includegraphics[width=0.35\textwidth]{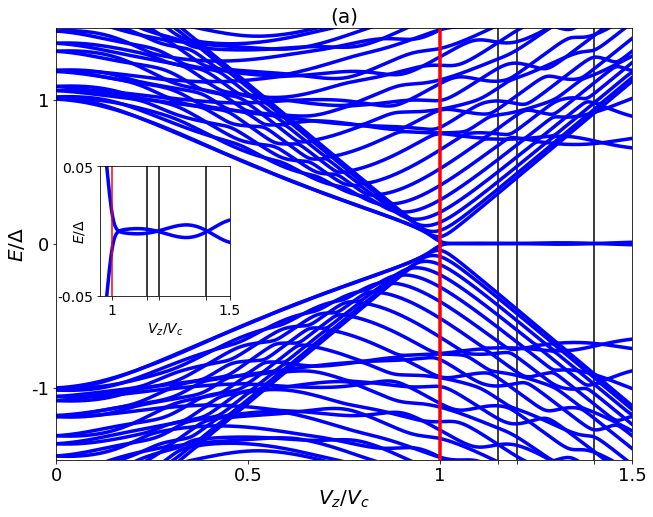}
	\hfill
	\centering
	\includegraphics[width=0.35\textwidth]{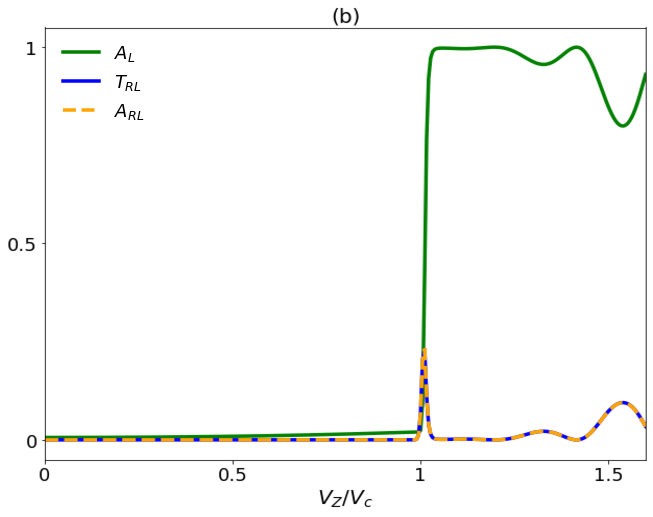}
	\caption{(a) Spectra of the pristine superconducting nanowire, described by equation (\ref{HTB}) as a function of the Zeeman field. In the inset, we show the Majorana oscillations. (b) Transmission probabilities for local Andreev reflection (green), $A_{L} = A_{R}$, direct transport (blue), $T_{RL} = T_{LR}$, and crossed Andreev reflection (orange), $A_{RL} = A_{LR}$, as a function of the Zeeman field. We note by comparing (a) and (b) that $T_{RL}$ and $A_{RL}$ follow the oscillations of the energy of the lowest mode. In addition, the conductance is dominated by LAR, which is enhanced by the emergence of the MZMs, in the regime of parameters considered. The parameters for the nanowire are detailed in Fig \ref{fig2}.}
	\label{spectraAndCoefficients}
\end{figure}

In Fig.~\ref{spectraAndCoefficients} (a), we show the spectrum of the pristine nanowire [superconducting section of the wire in Fig. \ref{fig1}(a)] as a function of $V_z/V_c$. The bulk topological phase transition occurs at $V_c = \sqrt{\Delta^2 + \mu^2}$ (vertical red line). For $V_z > V_c$, the Majorana zero modes located at the ends of the wire mediate a zero-bias conductance peak [Figs. \ref{fig1} (b), (c) and \ref{fig2} (c), (d) of the main text], which is dominated by LAR processes, as shown in Fig~\ref{spectraAndCoefficients} (b). As we discuss in the next section, the direct tunneling  ($T_{RL}$, blue line) and crossed Andreev reflection ($A_{RL}$, orange line) have the same amplitude and increase with the hybridization energy between the MZMs located at opposite ends of the wire. We observe that as the energy of the lowest mode (topological ABS) reaches a local maximum due to the oscillations seen in Fig. \ref{spectraAndCoefficients}(a)~\cite{Prada2012, DasSarmaSplitting, Rainis2013}, $T_{RL}$ and $A_{RL}$ also peak. These numerical results are consistent with the findings obtained via an effective model describing only the lowest modes coupled to metallic leads~\cite{splittingBeenakker}, as we show in the following.

\section{Low-energy effective Hamiltonian}
\label{lowenergyH}

In this section, we derive a low-energy effective Hamiltonian for the nanowire and its coupling to metallic leads. We start with a simple normal/superconductor (NS) junction with the superconductor in the topological phase. Projection onto the $\rm{N}+\rm{S}$ basis, described by the eigenvectors $\ket{N}$ and $\ket{S}$, results in the tunneling Hamiltonian
\begin{equation}
	\label{eq:HT1}
	H_T =  \sum_{\sigma = \uparrow, \downarrow} t_\sigma c^\dagger_{\sigma}(0) \psi_\sigma (0) +  t_\sigma^{*} \psi^\dagger_{\sigma}(0) c_\sigma (0),
\end{equation}
where $t_\sigma = \bra{N} H \ket{S}$, with $H$ the total Hamiltonian of the uncoupled system. Here, we use $c_\sigma(x) = \int dk c_{k, \sigma} e^{i k x}$ and $\psi_\sigma(x)$  to denote the annihilation operators on the normal and superconducting segments of the wire, respectively. Since the superconductor is in the topological phase, there is a Majorana mode localized near the domain wall at $x = 0$, whose operator is $\gamma = \int dx\left[ f \psi_\uparrow (x) + g \psi_\downarrow (x) + f^*  \psi_\uparrow^\dagger (x) + g^* \psi_\downarrow^\dagger (x)\right]$~\cite{flensberg2011non}. The spinor $\varphi^\dagger = \begin{pmatrix}f & g & g^* & -f^* \end{pmatrix}$ defining such an operator is obtained by solving the equation $\mathcal{H} \varphi (x) = 0$, where $\mathcal{H}$ is the Bogoliubov-de Gennes Hamiltonian of the superconducting nanowire in the basis $\{ \psi_\uparrow(x), \psi_\downarrow (x), \psi_\downarrow^\dagger (x), -\psi_\uparrow^\dagger (x) \}$. Additionally, we expand the operators $\psi_\sigma(x)$ in the basis of the excitations of the superconductor $\psi_{\uparrow ( \downarrow ) }(x) = f (g) \, \gamma + ... $~\cite{FuTeleportation}. We then make the assumption $\psi_{\uparrow ( \downarrow )} \sim f (g) \, \gamma$, with $|f|^2 + |g|^2 = 1$~\cite{PLeeSESAR}, which is a good approximation in the low-bias limit, where the main contribution to the conductance comes from the MZM. Using these operators in Eq.~\eqref{eq:HT1} we obtain 
\begin{equation}
    H_T = t \gamma \left[ f c_\uparrow (0) + g c_\downarrow (0) - f^* c_\uparrow^\dagger (0) - g^* c_\downarrow^\dagger (0) \right],
    \label{eq:HT2}
\end{equation}
with $t_\uparrow = t_\downarrow = t$ (unpolarized metal) real. 

It is possible to further simplify the Hamiltonian above  if one performs a unitary transformation such that $\Psi_1 = f c_\uparrow + g c_\downarrow$ and $\Psi_2 = -g^* c_\uparrow + f^*c_\downarrow$, which yields 
\begin{equation}
	H_T = t \gamma \left[ \Psi_1(0) - \Psi_1^\dagger (0) \right],
	\label{eq:HT3}
\end{equation}
i.e., the Majorana only couples to one type of electron in the metal~\cite{PLeeSESAR}.

\subsection{Local Andreev Reflection Coefficients}

Here we explicitly show that the spin-dependent local Andreev reflection coefficients shown in Fig. \ref{fig2}(a) of the main text can be traced back to the Majorana wavefunction. We use the scattering matrix formalism~\cite{splittingBeenakker, flensbergBCSCharge} and calculate the S-matrix for an NS junction,
\begin{equation} \label{smatrixGeneralEquation}
    S(E) = 1 - 2 \pi i \rho W^\dagger (E - \mathcal{H} + i \pi \rho W W^\dagger)^{-1} W,
\end{equation}
where $\mathcal{H}$ is the BdG Hamiltonian of the nanowire, which is zero for a semi-infinite wire in the low-energy regime, $\rho$ is the density of states of the normal lead, and $W = t \begin{pmatrix}f & g & -f^* & -g^* \end{pmatrix}$ in the basis $\{\Phi_{e, \uparrow}, \Phi_{e, \downarrow}, \Phi_{h, \uparrow}, \Phi_{h, \downarrow} \}$ of propagating electrons and holes in the lead. By substituting $W$ into Eq. (\ref{smatrixGeneralEquation}) we obtain

\begin{equation}
    S(E) = \begin{pmatrix} R^{ee} & A^{eh} \\ A^{he} & R^{hh} \end{pmatrix} = 1 - \frac{i\Gamma}{E + i \Gamma} \begin{pmatrix} |f|^2 & f^*g & -(f^*)^2 & -f^* g^* \\ g^*f & |g|^2 & -g^* f^* & -(g^*)^2 \\ -f^2 & -fg & |f|^2 & g^*f  \\ -fg & -g^2 & f^*g  & |g|^2 \end{pmatrix},
    \label{eq:Smatrix}
\end{equation}
with $\Gamma = 2 \pi \rho t^2$, $R^{ee(hh)}$ and $A^{he(eh)}$ the normal and Andreev reflection matrices, respectively. In particular, $A^{he}$ at $E = 0$ is given by

\begin{equation}
    A^{he} = \begin{pmatrix} a_{\uparrow \uparrow} & a_{\uparrow \downarrow} \\ a_{\downarrow \uparrow} & a_{\downarrow \downarrow} \end{pmatrix} = \begin{pmatrix}f^2 & fg \\ fg & g^2 \end{pmatrix}.
\end{equation}
Therefore, the Andreev components $A^L_{\sigma \sigma'} = |a_{\sigma \sigma'}|^2$ give us approximately the polarization of the Majorana bound state. As we increase the Zeeman field [Fig. \ref{fig2}(a) of the main text], the Majorana wavefunction starts to polarize in the spin-down direction. Hence as $|g|^2$ increases and $|f|^2$ decreases, $A^L_{\downarrow \downarrow}$ becomes the dominant process, which is already visible at $V_z/V_c \approx 1.40$. The transition between the spinful (small $V_z$) and the Kitaev chain (large $V_z$) regimes can be observed by measuring the individual Andreev reflection components.

One of the signatures of a MZM at $T=0$ is a quantized conductance of $2e^2/h$, even in the presence of two spin channels in the leads. This is distinct from the conductance for the BTK model~\cite{btk}, which can reach $4 e^2/h$. As shown in Eq.~\eqref{eq:HT3}, this is because the Majorana only couples to electrons with a certain spin polarization; from Eq.~\eqref{eq:Smatrix} we obtain that $R_L = \sum_{\sigma \sigma'} |r_{\sigma \sigma'}|^2 = A_L = \sum_{\sigma \sigma'} |a_{\sigma \sigma'}|^2 = 1$, with $r_{\sigma \sigma'}$ the coefficients of matrix $R^{ee}$.  This result can be understood by rewriting the S-matrix equation, relating incoming and outgoing electrons/holes, using the transformed basis $\{\psi_1$, $\psi_2, \psi_1^\dagger,\psi_2^\dagger\}$, i.e.,
\begin{equation}
	U^\dagger \Phi^{({\rm out})}(E) = U^\dagger S(E)  U U^\dagger\Phi^{({\rm in})}(E),
\end{equation}
 with $\Phi^{\rm (in/out)}(E) = \left(\Phi_{e, \uparrow} \ \Phi_{e, \downarrow} \ \Phi_{h, \uparrow} \ \Phi_{h, \downarrow}\right)^T$ and
\begin{equation}
    U^\dagger = \begin{pmatrix} \mathcal{U} & 0 \\ 0 & \mathcal{U}^* \end{pmatrix}, \,\,\,\, \mathcal{U} = \begin{pmatrix} f & g \\ -g^* & f^*
    \end{pmatrix}.
\end{equation}
The S-matrix equation then becomes
\begin{equation}
    \begin{pmatrix}
    \psi_1 (E) \\ \psi_2 (E) \\ \psi_1^\dagger (E) \\ \psi_2^\dagger (E)
    \end{pmatrix}^{({\rm out})} = \begin{pmatrix}
    \frac{E}{E + i \Gamma} &  0 & \frac{i\Gamma}{E+i\Gamma} & 0 \\ 
    0 & 1 & 0 & 0 \\ 
    \frac{i\Gamma}{E+i\Gamma}  & 0 & \frac{E}{E+i\Gamma} & 0 \\ 
    0 & 0 & 0 & 1
    \end{pmatrix} \begin{pmatrix}
    \psi_1 (E) \\  \psi_2 (E) \\ \psi_1^\dagger (E) \\ \psi_2^\dagger (E) 
    \end{pmatrix}^{({\rm in})}.
\end{equation}

As previously mentioned, there is no coupling between the subspaces of electrons $1$ and $2$, which ensures that all reflections are ``spin-conserving''. Moreover, incident type-2 electrons (holes) are always reflected as electrons (holes) with the same spin polarization, which means that this mode is effectively decoupled from the problem. Hence, the low-energy Hamiltonian of a Majorana wire coupled to external leads is spinless even outside of the strong Zeeman field regime. Finally, at $E = 0$ we have perfect Andreev reflection of type-1 electrons and holes. The Andreev reflection matrix in the new basis at $E = 0$ is

\begin{equation}
    \Tilde{A}^{he} = \begin{pmatrix} a_{11} & a_{12} \\ a_{21} & a_{22} \end{pmatrix} = \begin{pmatrix}1 & 0 \\ 0 & 0 \end{pmatrix}.
\end{equation}

\subsection{Conductance of a Majorana wire in a three-terminal device}
\label{Majoranacond}

We now show the derivation of the Majorana wire conductance for an NSN junction using the low-energy effective Hamiltonian $H_M = i \varepsilon_m \gamma_1 \gamma_2$, where $\varepsilon_m$ is the hybridization energy. As discussed in the previous section, the Majorana bound state couples to only one spin channel. This allows us to consider a spinless model for the leads without any loss of generality. 

Here we follow the derivation of the S-matrix presented in Ref.~\cite{splittingBeenakker}. In the basis $\begin{pmatrix}\Phi_{L, e} & \Phi_{R, e} & \Phi_{L, h} & \Phi_{R, h}  \end{pmatrix}^T$ of propagating electrons and holes in leads left $L$ and right $R$, we have

\begin{equation}
    S(E) = \begin{pmatrix}1 + A & A \\ A & 1 + A \end{pmatrix},
\end{equation}
where
\begin{equation} \label{matrixA}
\begin{split}
    &A = Z^{-1} \begin{pmatrix}i\Gamma_L (E + i \Gamma_R) & -\varepsilon_m \sqrt{\Gamma_L \Gamma_R} \\ \varepsilon_m \sqrt{\Gamma_L \Gamma_R} &  i\Gamma_R (E + i \Gamma_L)\end{pmatrix}, \\ 
    &Z = \varepsilon_m^2 - (E + i\Gamma_L)(E + i\Gamma_R),
\end{split}
\end{equation}
with $\Gamma_\alpha = 2\pi|t_\alpha|^2$. The matrix $A$ contains the crossed (off-diagonal terms) and local (diagonal) Andreev reflections. 

The zero-bias conductance in lead $\alpha$ is obtained via the current expression~\cite{flensbergBCSCharge, btk}
\begin{equation} \label{IL}
\begin{split}
    I_\alpha =& \frac{e}{h} \int dE \left(2 A_\alpha + T_{\beta \alpha} + A_{\beta \alpha} \right) \Tilde{f}(\mu_\alpha) +\\ &- \frac{e}{h} \int dE \left(T_{\alpha \beta} - A_{\beta \alpha} \right) \Tilde{f}(\mu_\beta),
\end{split}
\end{equation}
where $A_{\alpha}$, $T_{\beta \alpha}$, and $A_{\beta \alpha}$ are the probabilities of an incoming electron to be reflected as a hole in the same lead $\alpha$ and transmitted as an electron or as a hole to lead $\beta$, respectively, obtained from the S-matrix, $\Tilde{f}(\mu_\alpha) = f(E - \mu_\alpha) - f(E)$, $f(E) = \left[1 + e^{(\beta E)} \right]^{-1}$ is the Fermi function, and $\mu_\alpha$ is the chemical potential of lead $\alpha$ with respect to the chemical potential of the superconducting lead. In all conductance calculations, we consider $T = 0$ and small bias voltages. 

The local conductance in the left lead is given by 

\begin{equation} \label{GL}
\begin{split}
    G_{LL} =& \frac{e^2}{h}\left(2 A_L + T_{RL} + A_{RL} \right)|_{E \to 0} = \frac{2 e^2}{h} \frac{\Gamma_L \Gamma_R}{\varepsilon_m^2 + \Gamma_L \Gamma_R} \\ &=  \frac{2 e^2}{h} \frac{\Gamma^2 - \gamma_0^2}{\varepsilon_m^2 + \Gamma^2 - \gamma_0^2},
\end{split}
\end{equation}
where 
\begin{align}
	A_L &= |A_{11}|^2=\dfrac{\Gamma_L^2 \Gamma_R^2}{(\varepsilon_m^2 + \Gamma_L \Gamma_R)^2}\label{eq:AL}, \\
	T_{RL} &= A_{RL} = | A_{12}|^2=\dfrac{\varepsilon_m^2 \Gamma_L \Gamma_R}{(\varepsilon_m^2 + \Gamma_L \Gamma_R)^2}, \label{eq:TRL}
\end{align}
with $A_{ij}$ elements of the matrix \eqref{matrixA}, $\Gamma={\left(\Gamma_L + \Gamma_R\right)}/{2}$ and $\gamma_0={\left(\Gamma_L - \Gamma_R\right)}/{2}$. As the asymmetry grows ($\Gamma,\gamma_0 \rightarrow 1$), the conductance is suppressed, in agreement with Fig. \ref{fig3}. In the limit where one of the leads is decoupled ($\Gamma = \gamma_0$), $G_{LL}$ vanishes~\cite{flensbergMajoranaChain}. Note that the processes $T_{RL}$ and $A_{RL}$ qualitatively describe the behavior shown in Fig. \ref{spectraAndCoefficients}(b) for the full model. Whenever $E/\Delta$ crosses zero ($\varepsilon_m \rightarrow 0$) in \ref{spectraAndCoefficients}(a), $A_L$ is enhanced [Fig. \ref{spectraAndCoefficients}(b)]; $T_{RL}$ and $A_{RL}$, on the other hand, peak when $E/\Delta$ has a maximum ($\varepsilon_m$ increases), in agreement with Eq.~\eqref{eq:TRL}.

\section{Validity of the proposed protocol} \label{validityApp}

In order to assess the credibility and robustness of our protocol and possibly emulate experimental setups, we introduce additional nonuniformities to our superconducting nanowire segment and vary the disorder strength. We consider four different scenarios: (i) nontopological superconducting nanowire segment of length $\ell$, Figs.~\ref{nontopologicalmu:right} and \ref{nontopologicalmu:middle}, (ii) nonsuperconducting nanowire segment of length $\ell_N$, Figs.~\ref{nonsuperconducting}, (iii) asymmetric potential barriers, Fig.~\ref{asymmetricb}, and (iv) stronger disorder, i.e., $\sigma/\bar{\mu} > 1$~\cite{disorderDasSarma}, where $\sigma$ is the variance and $\bar{\mu}$ is the mean value, Fig.~\ref{disorder}. We find that, for the regime of parameters considered here ($t  = 102$, $\mu = 1$, $\mu_N = 0.2$, $\Delta = 0.5$, $\alpha = \alpha_R/2a = 3.5$, $U_L = 5$, $L_S = 2.5$, $L_N = 0.1$, and $L_B = 0.02$, unless otherwise specified; energies and lengths are in units of meV and $\mu$m, respectively), the protocol is valid as long as $\ell, \ell_N \lesssim 0.6 L_S$. For larger values of $\ell$ and $\ell_N$, it is still possible to observe a suppression of $G_{LL}$ as $U_R$ increases, but after an initial dip, the conductance attains a plateau, similarly to the quasi-Majorana case. For (iv), we find that, on average, the same behavior mentioned above for cases (i) and (ii) starts to occur within the range $2.5 \lesssim \sigma/\bar{\mu} \lesssim 3$. For case (iii), we consider the right side barrier with lengths $L_{B,R} = 0.04$ and $0.08~\rm{\mu m}$ (for the left barrier $L_{B,L}=0.02~{\mu}$m). We find that the correlation $G_{LL} = G_{RR}$ holds, and the conductance suppression due to the increase of $U_R$ is more pronounced for larger $L_{B, R}$.

\subsection{Nontopological superconducting nanowire segment}

\begin{figure}[htb!]
	\centering
	\includegraphics[width=0.3\textwidth, keepaspectratio]{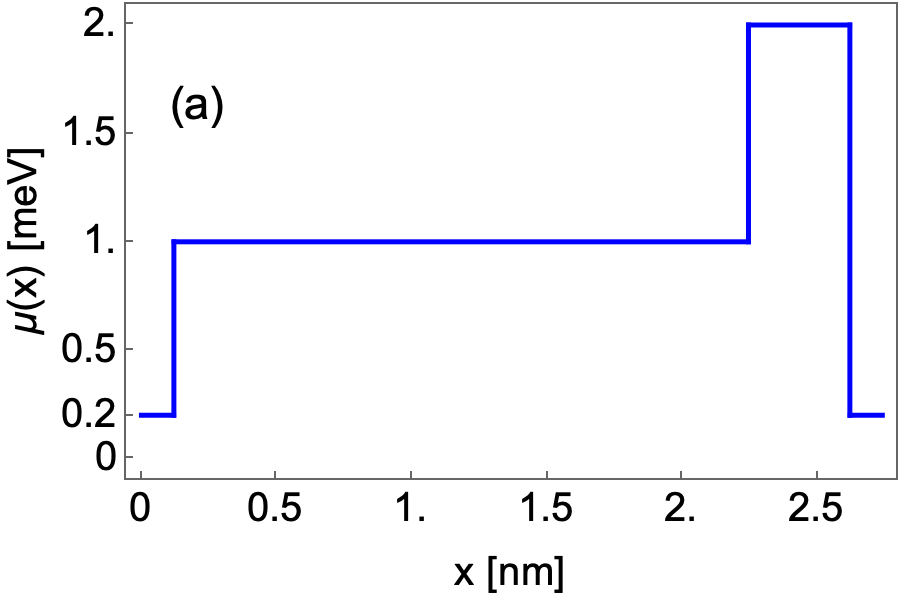}
	\centering
	\includegraphics[width=0.235\textwidth]{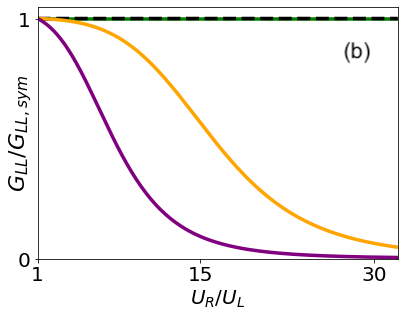}
	\centering
	\includegraphics[width=0.235\textwidth]{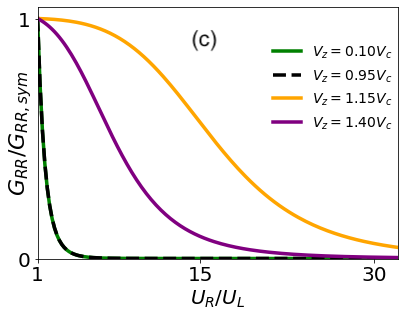}
	\centering
	\includegraphics[width=0.3\textwidth, keepaspectratio]{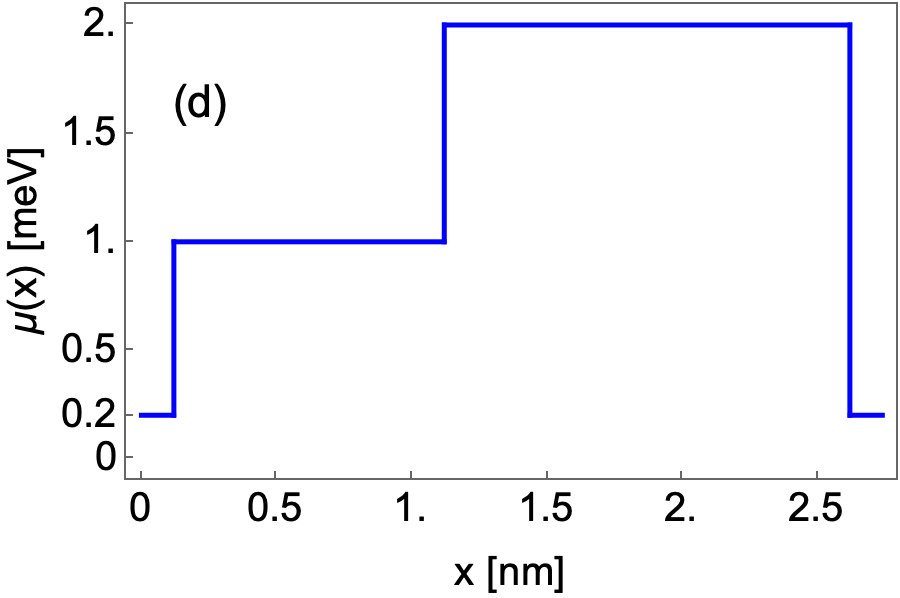}
	\centering
	\includegraphics[width=0.235\textwidth]{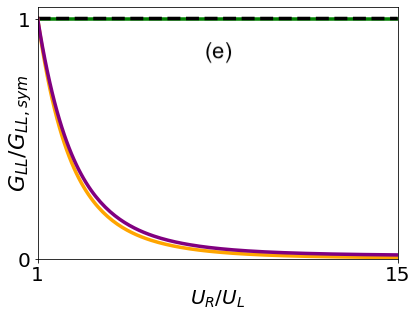}
	\centering
	\includegraphics[width=0.235\textwidth]{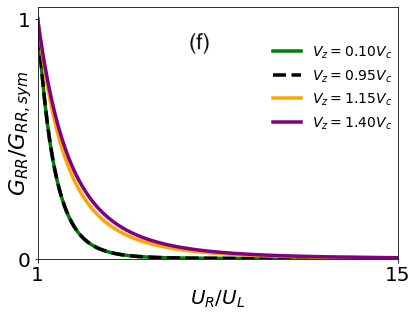}
	\caption{Chemical potential profile for a trivial region ($\tilde{\mu}=2$~meV) of length (a) $\ell=0.375~\rm{nm}$ and (d) $\ell=1.5~\rm{nm}$.  $G_{LL}$ and $G_{RR}$ as functions of $U_R/U_L$ for (b), (c) the profile in (a), and (e), (f) the profile in (d). }
	\label{nontopologicalmu:right}
\end{figure}

\begin{figure}[htb!]
	\centering
	\includegraphics[width=0.3\textwidth, keepaspectratio]{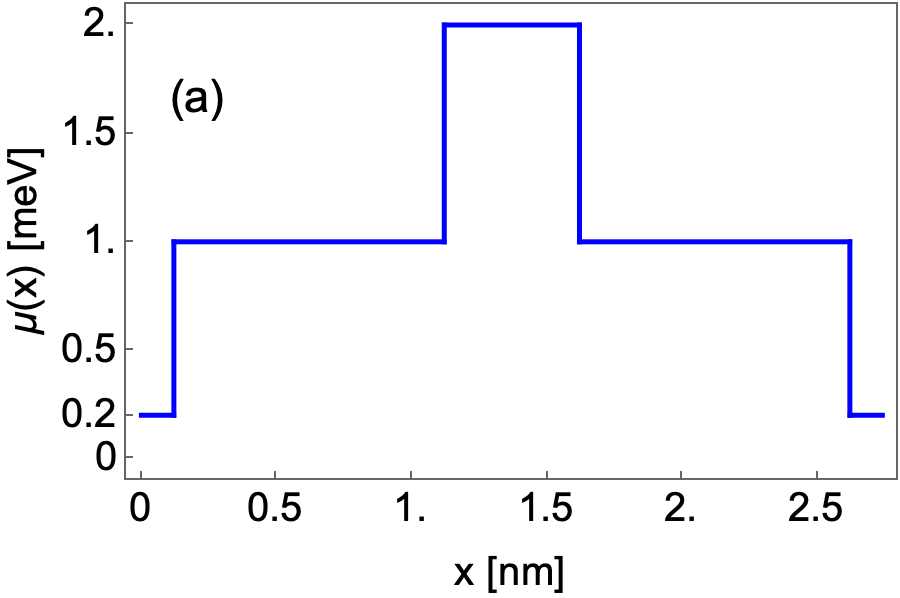}
	\centering
	\includegraphics[width=0.235\textwidth]{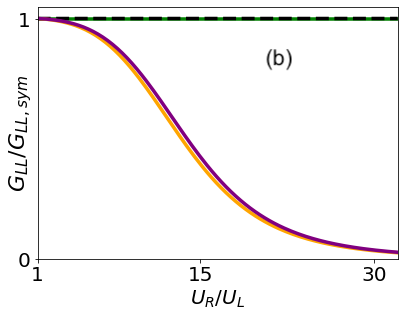}
	\centering
	\includegraphics[width=0.235\textwidth]{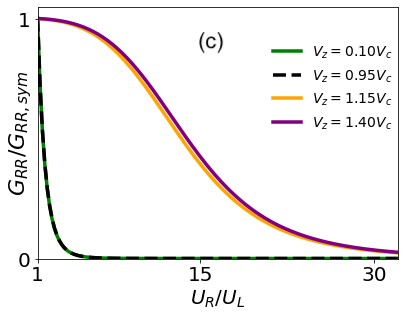}
	\centering
	\includegraphics[width=0.3\textwidth, keepaspectratio]{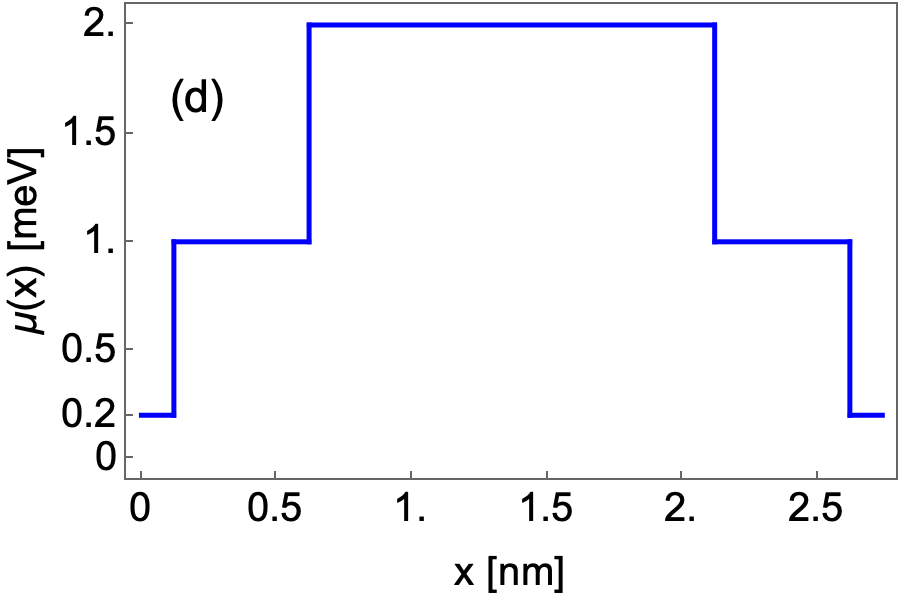}
	\centering
	\includegraphics[width=0.235\textwidth]{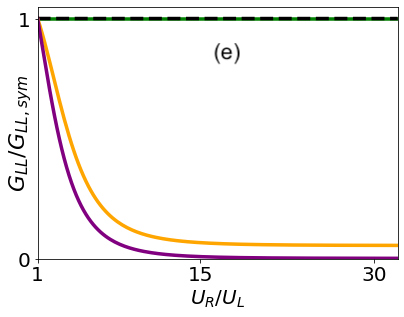}
	\centering
	\includegraphics[width=0.235\textwidth]{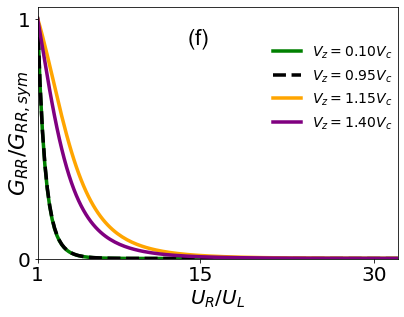}
	\caption{Chemical potential profile for a trivial region ($\tilde{\mu}=2$~ meV) of length (a) $\ell=0.5~\rm{nm}$ and (d) $\ell=1.5~\rm{nm}$.  $G_{LL}$ and $G_{RR}$ as functions of $U_R/U_L$ for (b), (c) the profile in (a), and (e), (f) the profile in (d). }
	\label{nontopologicalmu:middle}
\end{figure}

Here we consider another nonuniform profile for $\mu_S$, the chemical potential of the SC section of the nanowire. We choose a segment of length $l$, either located at one of the ends [Fig.~\ref{nontopologicalmu:right} (a) and (d)] or in the middle of the SC nanowire [Fig.~\ref{nontopologicalmu:middle}(a) and (d)], whose chemical potential $\tilde{\mu}=2$~meV is such that this segment is in the trivial regime.

In Figs.~\ref{nontopologicalmu:right} (a) and (d), we show $\mu_S(x)$ as a function of $x$ for $\ell=0.375$ and $1.5~\rm{\mu m}$ $(0.6 L_S)$, respectively. We note that $G_{LL} \approx G_{RR}$ when $V_z>V_c$, signaling the topological phase. For larger values of $\ell$, our protocol breaks down since $G_{LL}$ and $G_{RR}$ are suppressed differently, see some examples below.  In this case, we can no longer distinguish between the trivial and nontrivial regimes. 

Similarly in Figs.~\ref{nontopologicalmu:middle} (a) and (d), we show $\mu_S(x)$ for a segment $\ell=0.5$ and $1.5~\rm{nm}$, respectively, located in the middle of the SC nanowire. The curves for $G_{LL}$ and $G_{RR}$, especially in (e) and (f), start to differ from each other and our protocol can no longer identify the topological phase. 

\subsection{Nonsuperconducting nanowire segment}

\begin{figure}[htb!]
	\centering
	\includegraphics[width=0.3\textwidth, keepaspectratio]{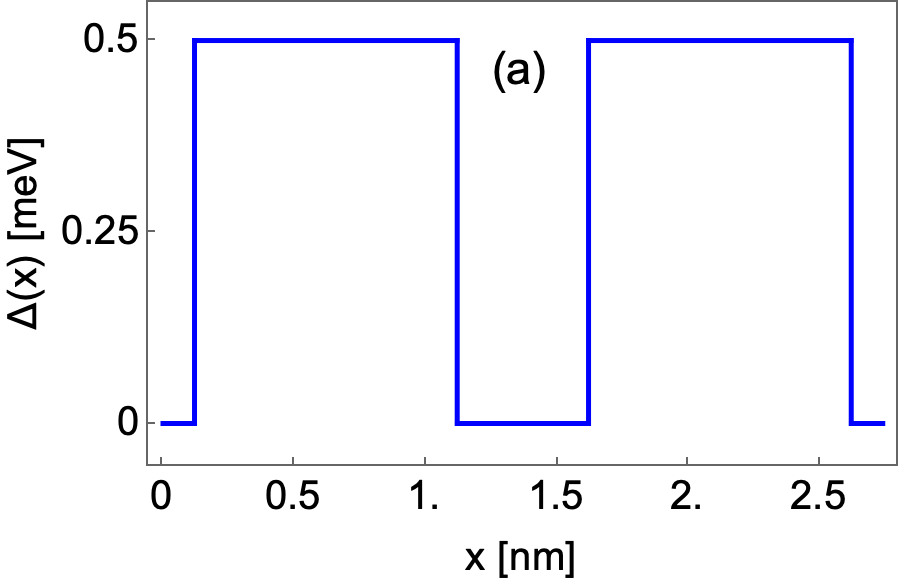}
	\centering
	\includegraphics[width=0.235\textwidth]{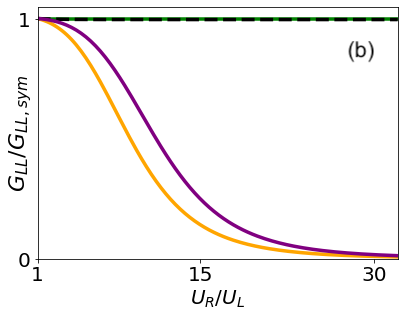}
	\centering
	\includegraphics[width=0.235\textwidth]{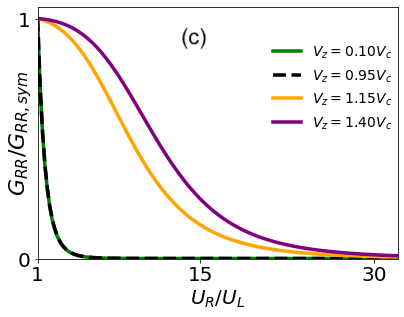}
	\centering
	\includegraphics[width=0.3\textwidth, keepaspectratio]{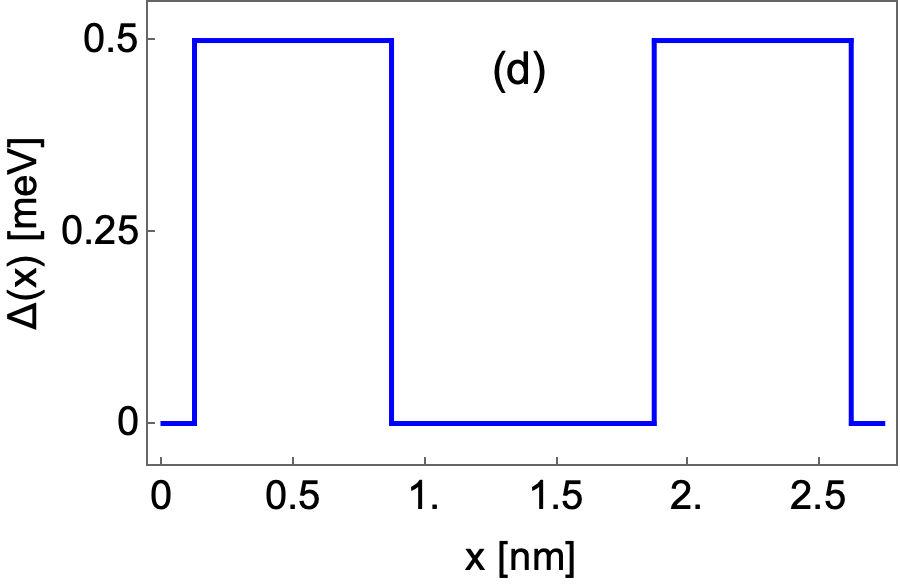}
	\centering
	\includegraphics[width=0.235\textwidth]{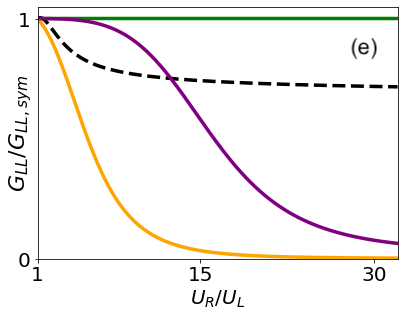}
	\centering
	\includegraphics[width=0.235\textwidth]{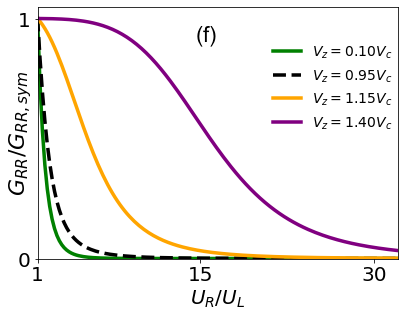}
	\caption{Superconducting pairing potential profile for a trivial region ($\Delta=0$) of length (a) $\ell=0.5~\rm{nm}$ and (d) $\ell=1.0~\rm{nm}$.  $G_{LL}$ and $G_{RR}$ as functions of $U_R/U_L$ for (b), (c) the profile in (a), and (e), (f) the profile in (d). }
	\label{nonsuperconducting}
\end{figure}

Here we modify the profile of the superconducting pairing potential $\Delta$ and introduce a segment, $\ell_N$, in the middle of the SC nanowire for which $\Delta=0$, Fig.~\ref{nonsuperconducting}.

\subsection{Asymmetric potential barriers}

\begin{figure}[H]
	\centering
	\includegraphics[width=0.3\textwidth, keepaspectratio]{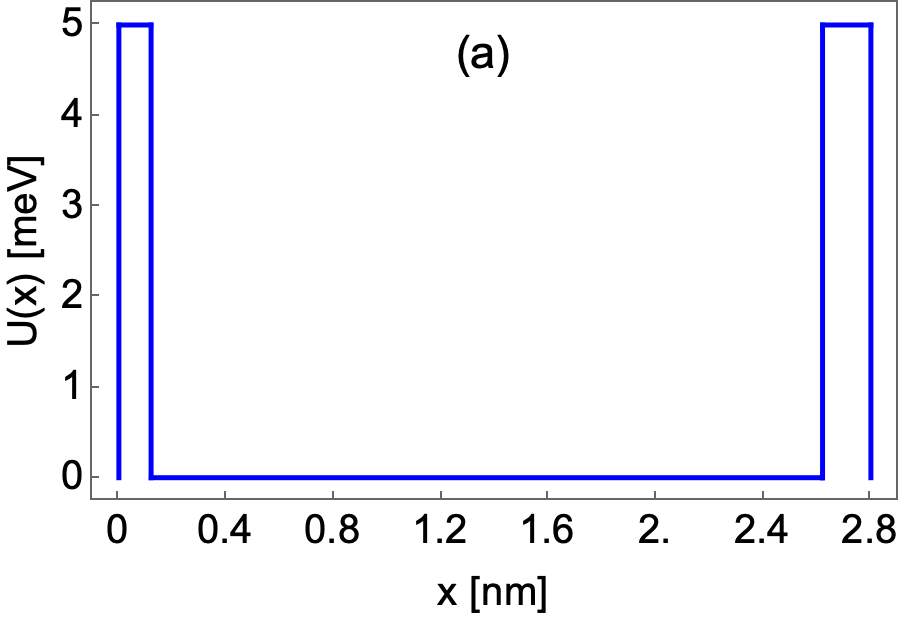}
	\centering
	\includegraphics[width=0.235\textwidth]{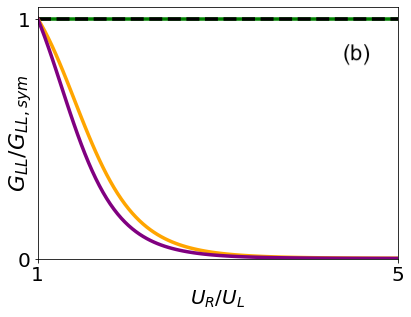}
	\centering
	\includegraphics[width=0.235\textwidth]{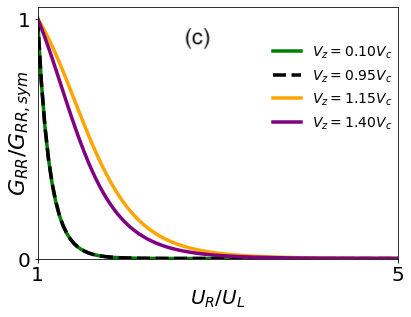}
		\caption{(a) Potential barrier profile: $L_{B,L}=0.02~\rm{nm}$ (left) and $L_{B,R}=0.08~\rm{nm}$. (b) $G_{LL}$ and (c) $G_{RR}$ as functions of $U_R/U_L$.}
	\label{asymmetricb}
\end{figure}

In Fig.~\ref{asymmetricb}(a), we show the potential barrier profile for $L_{B,R}=4L_{B,L}$. The conductances $G_{LL}$ and $G_{RR}$, Fig~\ref{asymmetricb} (b) and (c), respectively, still behave similarly. We observe that they are suppressed more rapidly when $L_{B,R}$ increases (not shown). 

\subsection{Strong disorder}

\begin{figure}[htb!]
\centering
\includegraphics[width=0.4\textwidth]{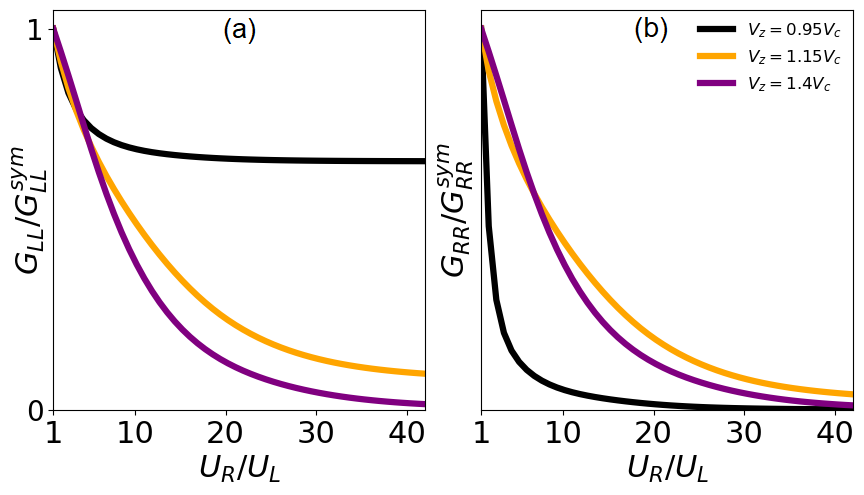}
\centering
\includegraphics[width=0.4\textwidth]{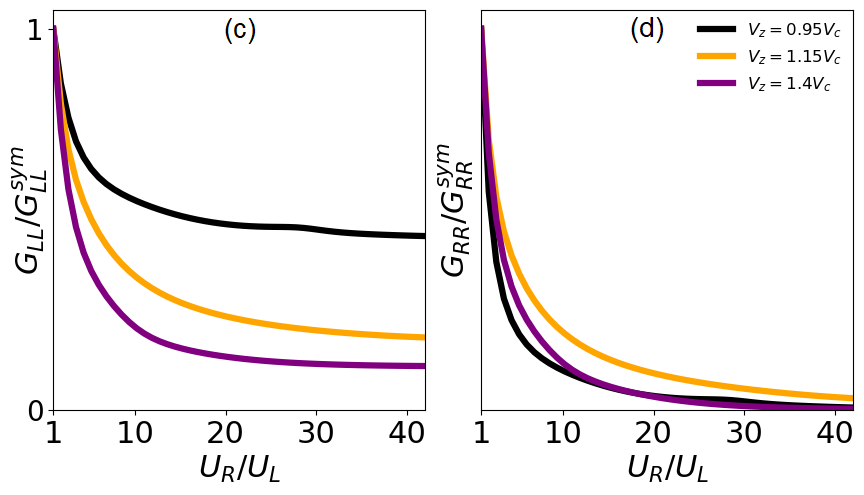}
\caption{$G_{LL}$ and $G_{RR}$ as functions of $U_R/U_L$ for (a), (b) $\sigma/\bar{\mu}=1.5$ and (c), (d)  $\sigma/\bar{\mu}=3.0$.}
\label{disorder}
\end{figure}

Here we show $G_{LL}$ and $G_{RR}$ as functions of $U_R/U_L$ for two different strengths of disorder (averaged over $100$ realizations): Figs.~\ref{disorder} (a) and (b) $\sigma/\bar{\mu}=1.5$ ($W=5.2\Delta$) and (c) and (d) $\sigma/\bar{\mu}=3.0$ ($W=10.45\Delta$). As previously mentioned, for $\sigma/\bar{\mu}=3.0$ our protocol is no longer valid, since $G_{LL}$ and $G_{RR}$ are suppressed differently. 

\subsection{Emergence of trivial extended ABSs}

Fluctuations in the chemical potential generate a chain of Majorana modes for large enough magnetic fields ($V_z < V_c$), as many smaller trivial/topological junctions are formed within the wire \cite{flensbergMajoranaChain}. The hybridization between these modes may give rise to extended Andreev bound states. These ABSs might induce a considerable dependence of $G_{LL}$ on $U_R$ since they couple to both leads. In general, short disordered wires are problematic for experiments involving Majoranas because extended trivial states can mimic features commonly associated with the emergence of the topological phase, such as the closing and reopening of the gap \cite{NonLocalLoss}. In our simulations, we have verified that nanowires of $1.5~\mu$m or shorter with $W = 3.5~\Delta$ are undesirable, as the emergence of such states is quite common. On the other hand, by increasing the length of the wire to $2~\mu$m, for the same disorder strength, we obtain significantly better results. 
\begin{figure}[htb!]
	\centering
	\includegraphics[width=0.4\textwidth]{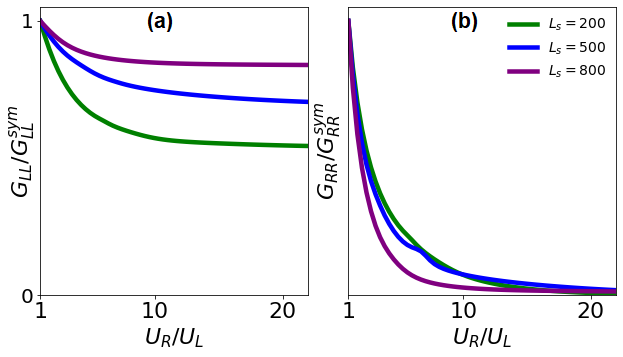}
	\centering
	\includegraphics[width=0.4\textwidth]{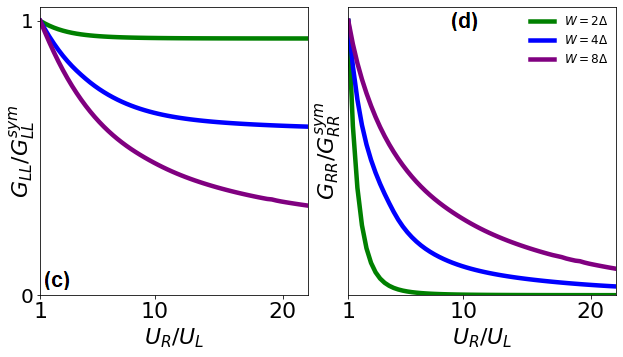}
	\caption{ Disordered nanowire (averaged over $1000$ realizations) conductances $G_{LL}$ and $G_{RR}$ in the trivial phase ($V_z = 0.95V_c$) as functions of $U_R/U_L$ for different (a), (b) lengths of the superconducting section $L_s = 1, 2.5, 4~\mu$m with $W = 3.5 \Delta$, and (c), (d) disorder strengths $W = 2\Delta, 4\Delta, 8\Delta$ for $L_s = 2.5~\mu$m. }
	\label{EmergenceExtendedABS}
\end{figure}

To systematically study the emergence of trivial extended ABSs in disordered wires, we perform averages on the local conductances as functions of $U_R/U_L$ for $1000$ disorder realizations at $T = 0$ and $V_z = 0.95 V_c$. In Fig. \ref{EmergenceExtendedABS}(a), we fix $W = 3.5\Delta$ and vary the length of the wire, $L_s$. We observe that as $L_s$ increases, the dependence of $G_{LL}$ on $U_R/U_L$ on average decreases (cp., for example, the green and purple curves). Therefore, in longer wires, the probability of an extended ABS emerging is lower. As discussed in Ref.~\cite{flensbergMajoranaChain}, weak links in such chains of Majorana modes effectively decouple the two sides. Hence the probability of the chain being randomly broken due to a weak link grows with the length of the wire, thus reducing, on average, the emergence of extended ABSs. Using the suppression of $G_{LL}$ for $U_R >> U_L$ as an indication of the presence of extended ABS, we have verified that by increasing the disorder strength $W$ ($L_s = 2.5 \mu$m fixed) there is a corresponding increase in the probability of such states emerging. As we diminish $W$, and in particular for $W \leq 2\Delta$, the conductances approach the results for the pristine case, shown in Figs. 2(c)-(d) of the main text.

\subsection{Symmetric responses to variations of the tunnel gate voltages}

\begin{figure}[htb!]
	\centering
	\includegraphics[width=0.285\textwidth]{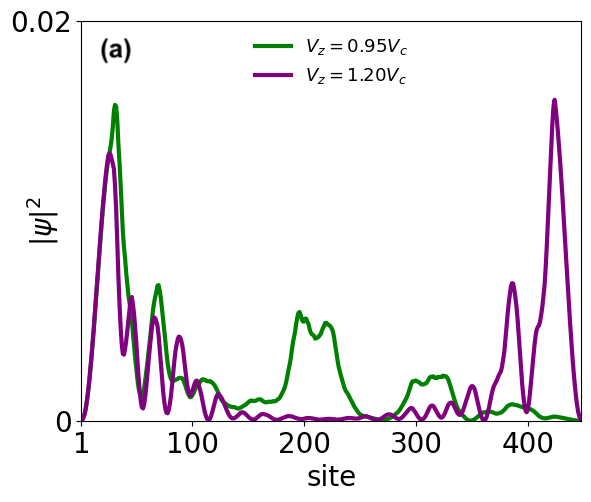}
	\centering
	\includegraphics[width=0.45\textwidth]{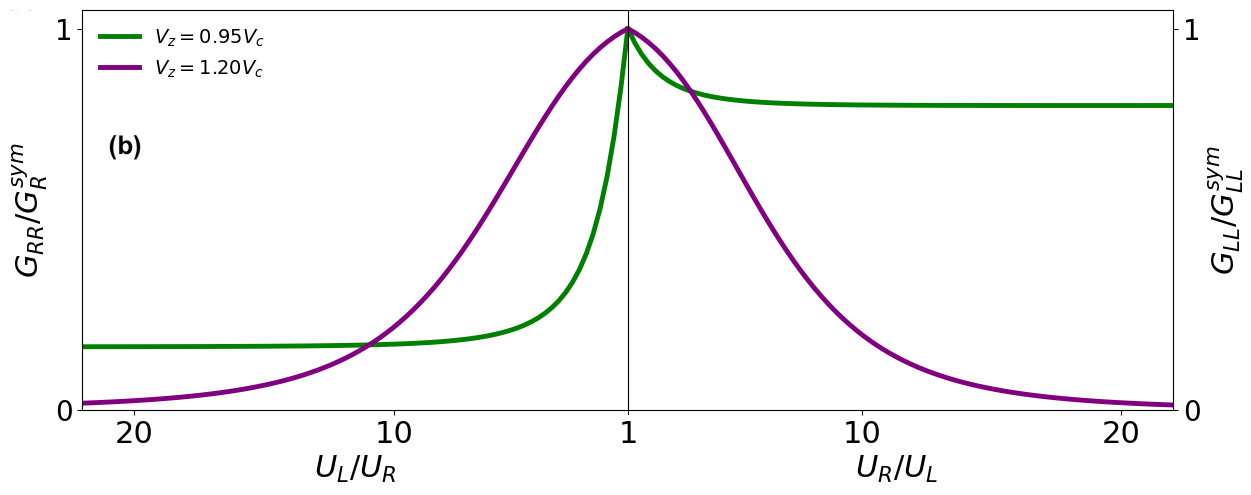}
	\caption{(a) Probability density of the lowest-energy modes of a disordered (single realization)  wire with $W = 3.5 \Delta$, $L_s = 400$, and $U_L = U_R = 5~$meV, in the trivial (green) and topological (purple) phases. (b) $G_{LL}$ and $G_{RR}$ as functions of $U_R/U_L$ and $U_L/U_R$, respectively. }
	\label{LeftvsRight symmetry}
\end{figure}

Here we provide yet an additional way of distinguishing topological from extended trivial Andreev bound states. We analyze how the local conductances, $G_{LL}$ and $G_{RR}$, respond to variations on the gate that controls the coupling to the lead on the opposite side of the wire, $U_R$ and $U_L$, respectively. If the response is symmetric, i.e., if $G_{LL}$ vs. $U_R$ has the same behavior as $G_{RR}$ vs. $U_L$, the system is in the topological regime, otherwise, we have a trivial phase. We first note that the \textit{nonlocality} of the local conductances comes from a single state coupling to both leads. In singular disorder realizations, an extended trivial ABS might be able to present a large dependence of $G_{LL}$ with respect to $U_R$, for instance. In general, however, the wavefunction of trivial states will not be symmetrically distributed along the wire, resulting in different couplings to the leads even when $U_R = U_L$, see for instance, the green curve for $V_z = 0.95 V_c$ in Fig.~\ref{LeftvsRight symmetry}(a). Although the trivial ABS extends throughout the entire wire, thus connecting to both leads, the coupling to the right lead is much smaller than the one to the left lead, i.e. $\Gamma_R << \Gamma_L$, for $U_R = U_L$. Hence, $G_{LL}$ is not significantly suppressed by the increase of $U_R$, as the initial coupling $\Gamma_R$ is already small. On the other hand, $\Gamma_L$ is large, since the wavefunction is more localized on the left end of the wire. In this case, $G_{RR}$ is significantly more suppressed by the increase of $U_L$. This behavior is corroborated by our simulations of the conductances $G_{LL}$ and $G_{RR}$ as functions of $U_{R}$ and $U_L$, respectively,  see green curves in Fig. \ref{LeftvsRight symmetry}(b). In contrast, topological ABSs that arise from the hybridization between MZMs are, in general, located at both ends of the wire, even in the presence of (up to moderate) disorder. This approximately symmetric distribution of the wavefunction, purple curve in Fig. \ref{LeftvsRight symmetry}(a), leads to $\Gamma_L \sim \Gamma_R$, when $U_R = U_L$, resulting in symmetric responses $G_{LL}(U_R) \sim G_{RR}(U_L)$, see corresponding purple curve in Fig. \ref{LeftvsRight symmetry}(b).

\subsection{Temperature effects in the trivial phase}

\begin{figure}[htb!]
\centering
\includegraphics[width=0.4\textwidth]{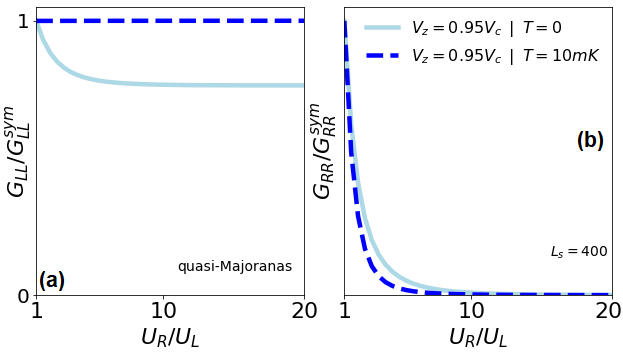}
\centering
\includegraphics[width=0.4\textwidth]{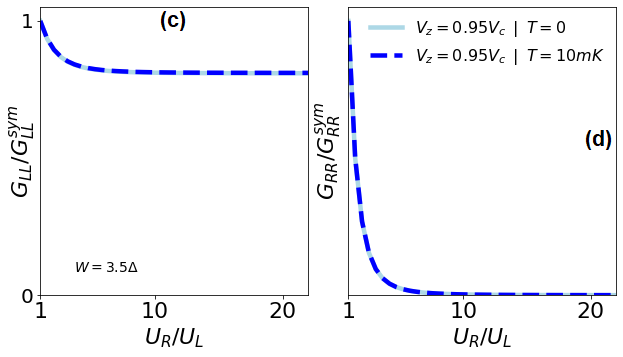}
\caption{$G_{LL}$ and $G_{RR}$ as functions of $U_R/U_L$ for (a), (b) the quasi-Majorana profile and (c), (d) a disordered nanowire with $W = 3.5 \Delta$. We set $L_s = 400$ for all cases.}
\label{Disorder and QM finite T}
\end{figure}

In the main text, we have shown that the minimum value of $G_{LL}$ [plateaus in Fig. 5(a) of the main text] as $U_R$ increases is limited by the ratio $\varepsilon_m/k_BT$. Here we extend our analysis to the trivial phase. We observe that both quasi-Majoranas and disordered driven states respond differently to finite temperatures. For quasi-Majoranas, see Figs. \ref{Disorder and QM finite T} (a) and (b), we find that the initially small suppression of $G_{LL}$ at $T=0$ (light blue line) is entirely lost even at $T=10$~mK (dashed blue line). This is in contrast to the results shown in Fig. 5(a) of the main text, in which the topological state presents approximately the same results for $T = 0$ and $10$~mK. Again, temperature effects are beneficial to distinguishing quasi-Majoranas from MZMs. The conductance $G_{LL} (U_R)$ for disorder-driven states tends to show a much smaller dependence on temperature, as shown in Figs. \ref{Disorder and QM finite T} (c) and (d).

\section{Non-Hermitian topology and Exceptional points} \label{nonHermitianSection}

In this section, we explore the physics of Exceptional Points (EPs)~\cite{heiss2012physics} to complement our analysis of the LDOS via Green functions. First, we project the Hamiltonian of the Majorana wire in the low energy subspace $H_M$ using the BdG formalism, and add the self-energy $\Sigma_\alpha =- i\Gamma_\alpha$ to it, similarly to what was done in Ref.~\cite{aguadoNHT}. The non-Hermitian Hamiltonian reads
\begin{figure}[htb!]
	\centering
	\includegraphics[width=0.35\textwidth]{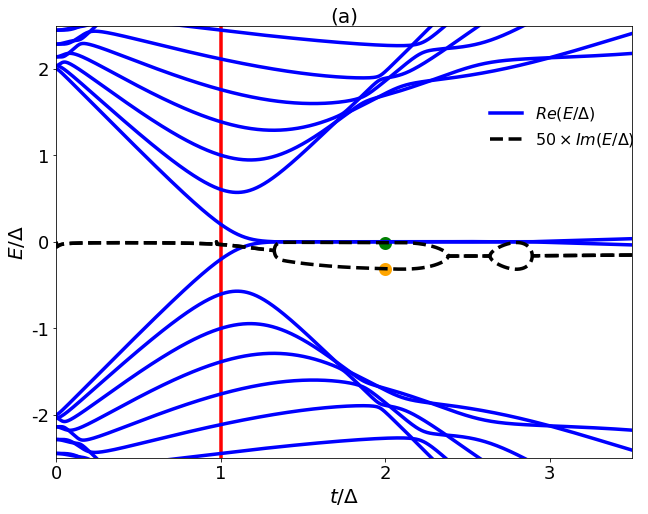}
	\hfill
	\centering
	\includegraphics[width=0.35\textwidth]{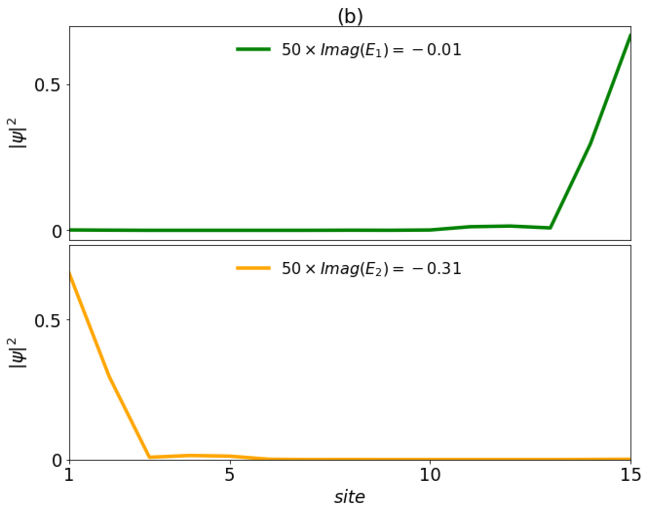}
	\caption{(a) Spectra of a non-Hermitian Hamiltonian of a Kitaev chain coupled to external leads $H_{eff} = H_{\rm K} + \Sigma$, using $N = 15$, $\mu/\Delta = 2$, $\Gamma_L = 5 \varepsilon_m$, and $\nu = 0.95$. Note that whenever the energy is smaller than a certain threshold, there is a bifurcation of the imaginary part of the lowest mode and its particle-hole partner ($E_1$ and $E_2$) (b) Probability densities for the two lowest modes for $t/\Delta = 2$ indicated by the green and orange dots in (a). The left Majorana (orange) leaks to the lead due to a larger value of its imaginary part, while the right Majorana (green) has a significantly larger lifetime.}
	\label{NonHermitianFig}
\end{figure}

\begin{equation} \label{heffMajorana}
	H_{\rm eff} = \frac{1}{2}\begin{pmatrix} \gamma_1 & \gamma_2 \end{pmatrix} \begin{pmatrix} -i\Gamma_L & i \varepsilon_m \\ -i \varepsilon_m & -i \Gamma_R \end{pmatrix} \begin{pmatrix}\gamma_1 \\ \gamma_2 \end{pmatrix}.
\end{equation}
It is convenient to rewrite the coupling to the leads in terms of symmetric $\Gamma$ and anti-symmetric $\gamma_0$ parts. Note also that the lead-asymmetry defined in the main text is $\nu = \gamma_0/\Gamma$. The BdG Hamiltonian then becomes
\begin{equation}
	\mathcal{H}_{\rm eff} = \begin{pmatrix}
		-i \Gamma -i\gamma_0 & i \varepsilon_m \\ -i \varepsilon_m & -i \Gamma + i \gamma_0
	\end{pmatrix},
\end{equation}
whose eigenenergies and eigenvectors are 
\begin{equation}
	\varepsilon_\pm = -i \Gamma \pm \sqrt{\varepsilon_m^2 - \gamma_0^2}
\end{equation}
and
\begin{equation}
	\ket{\phi_\pm} = \begin{pmatrix}
		1 \\ a_\pm
	\end{pmatrix}, \,\,\,   a_\pm = \frac{\gamma_0 \pm \sqrt{\gamma_0^2 - \varepsilon_m^2}}{\varepsilon_m}.
\end{equation}

The EPs arise when the eigenvectors of the non-Hermitian Hamiltonian coalesce into one, in this case, $\ket{\phi_+} = \ket{\phi_-}$, i.e. $a_+ = a_-$. It is easy to see that this occurs when $\gamma_0 = \pm \varepsilon_m$, and therefore there are two EPs. For $\gamma_0^2 > \varepsilon_m^2$, the real part of the eigenenergy goes to zero, while the imaginary part bifurcates into $\Gamma_\pm = - (\Gamma \pm \sqrt{\gamma_0^2 - \varepsilon_m^2})$. This result can be understood as one of the Majorana modes leaking into one of the leads ($\Gamma_+$) while the other has an increased lifetime ($\Gamma_-$). Interestingly, the leakage does not depend only on the coupling of the Majorana to its respective lead. For example, if we fix $\Gamma_L$ and \textit{decrease} $\Gamma_R$, the left Majorana starts to leak \textit{more} to the left lead as $\gamma_0$ increases. This is a consequence of the nonlocality of the MZMs' wavefunctions. In the main text, the same conclusions are obtained by solving the Green functions; for large $\nu$, a gap emerges in the LDOS on the side that is more strongly coupled to the lead, as shown in Fig. \ref{fig3}.

To illustrate the behavior above, we numerically obtain the spectrum of the open system using the full Kitaev chain, described by the Hamiltonian
\begin{equation}
	H_{\rm K} = -\mu \sum_{j=1}^{N} c_j^\dagger c_j + \sum_{j=1}^{N-1}\left(\Delta c^\dagger_j c^\dagger_{j+1} - t c^\dagger_j c_{j+1} + \rm{H.c.}\right),
\end{equation}
for  $N = 15$, $\mu/\Delta = 2$, $\Gamma_L = 5 \varepsilon_m$, $\gamma_0=190 \varepsilon_m$, and $\nu = 0.95$. The hybridization energy $\varepsilon_m$ (energy of the lowest mode), used as a reference here, was obtained at $t/\Delta=2.5$ for the closed system.  In Fig.~\ref{NonHermitianFig}(a) we show $E/\Delta$ as a function of the hopping parameter $t/\Delta$. We note that whenever the energy, which oscillates as $t/\Delta$ increases, is smaller than $\gamma_0$, there is a bifurcation in the imaginary part of the eigenenergy while its real part goes to zero. In Fig.~\ref{NonHermitianFig}(b), we show the probability densities for the two lowest modes for $t/\Delta = 2$, indicated in Fig.~\ref{NonHermitianFig}(a) by the green and orange points. We note that the imaginary part of the energy of the Majorana on the left (orange) is larger than the one located on the right (green). This means that it leaks faster to the lead, in agreement with the result (gap in the LDOS) in Fig. \ref{fig3}. We point out that if we fix $\Gamma_R$ and vary $\Gamma_L$ instead, the behavior of left and right Majoranas invert in terms of the imaginary part of the energy, i.e., the system is symmetric with respect to varying $\Gamma_R$ ($\Gamma_L$) and keeping $\Gamma_L$ ($\Gamma_R$) fixed. In conclusion, if we keep the parameters constant, and therefore $\varepsilon_m$, the emergence of the exceptional points can be controlled by the asymmetry in the coupling to the left and right leads $\nu$.

\section{Green's function and conductance of the Majorana-based transistor} \label{AppTransistor}

The complete Hamiltonian is $H = H_0 + H_T$, where $H_0$ is the system's Hamiltonian

\begin{equation}
	H_0 = i \varepsilon_m \gamma_1 \gamma_2 + \epsilon_d c^\dagger_d c_d + t_0 \left(c_d^\dagger - c_d \right) \gamma_1
\end{equation}
and $H_T$ represents the coupling to the normal leads

\begin{equation}
	H_T =  \sum_{k, i = 1, 2} t_{k,i} ( c^\dagger_d d_{k, i} + H.c.) + \sum_k t_{k, R} (d_{k, R}^\dagger - d_{k, R} ) \gamma_2.
\end{equation} 
First, we write $H_0$ in the BdG formalism.

\begin{equation}
	H_0 = \frac{1}{2} \Psi^\dagger \mathcal{H}_{0} \Psi, \,\,\, \Psi = \begin{pmatrix}c_d & c_d^\dagger & \gamma_1 & \gamma_2 \end{pmatrix}^T,
\end{equation}
where

\begin{equation} \label{H0BdG}
	\mathcal{H}_{0} = \begin{pmatrix}\epsilon_d & 0 & t_0 & 0 \\ 
		0 & -\epsilon_d & -t_0 & 0 \\ 
		t_0 & -t_0 & 0 & i \epsilon_m \\
		0 & 0 & -i \epsilon_m & 0 \end{pmatrix}.
\end{equation}
To obtain the self-energies, we use the expression $\Sigma_i = \sum_k \mathcal{H}_{Ti}^\dagger g_i \mathcal{H}_{Ti}$, where $g_i$ are the GFs of the isolated leads, and $\mathcal{H}_{Ti}$ the matrices coupling the system to the leads, which are given by

\begin{equation}
	\mathcal{H}_{TR} = \begin{pmatrix}0 & 0 & 0 & t_R \\  0 & 0 & 0 & -t_R \end{pmatrix}, \,\,\, \mathcal{H}_{T1(2)} = \begin{pmatrix}t_{1(2)} & 0 & 0 & 0 \\  0 & -t_{1(2)} & 0 & 0 \end{pmatrix}.
\end{equation}
By making the assumption that the self-energies do not depend on the energy (wide-band limit), and $\Gamma_1 = \Gamma_2 = \Gamma_L$, we obtain $\Sigma = \Sigma_1 + \Sigma_2 + \Sigma_R$

\begin{equation}
	\Sigma = \begin{pmatrix}-i\Gamma_L & 0 & 0 & 0 \\
		0 & -i\Gamma_L & 0 & 0 \\
		0 & 0 & 0 & 0 \\ 0 & 0 & 0 & -i \Gamma_R \end{pmatrix}.
\end{equation}
Finally, the GF is

\begin{equation}
	\mathcal{G}^r(E) = (E - \mathcal{H}_{0} - \Sigma)^{-1},
\end{equation}

\begin{equation} \label{GFTotal}
	\begin{split}
		\mathcal{G}^r &= \begin{pmatrix}E + i\Gamma_L-\epsilon_d & 0 & -t_0 & 0 \\ 
			0 & E + i\Gamma_L + \epsilon_d & t_0 & 0 \\ 
			-t_0 & t_0 & E & -i \epsilon_m \\
			0 & 0 & i \epsilon_m & E + i \Gamma_R \end{pmatrix}^{-1} \\
		&= \begin{pmatrix}\mathcal{G}^r_{\rm{dot}} & \Tilde{F}
			\\ \Tilde{F} & \mathcal{G}^r_{\rm{MW}} \end{pmatrix},
	\end{split}
\end{equation}
where we identify the first $2 \times 2$ block as the GF of the dot.

\begin{equation}
	\mathcal{G}^R_{\rm{dot}} = \begin{pmatrix}\mathcal{G}_e & F \\ F & \mathcal{G}_h \end{pmatrix},
\end{equation}
where, considering low energies ($E^2 \sim 0$),
\begin{subequations}
	\begin{equation}
		F =  -t_0^2 (E + i \Gamma_R) \tilde{Z}^{-1},
	\end{equation}
	\begin{equation}
		\mathcal{G}_{e(h)} = \tilde{Z}^{-1}\left[-\varepsilon_m^2 (\pm \epsilon_d + i \Gamma_L + E) + E ( \pm  i \epsilon_d \Gamma_R  - \Gamma_L \Gamma_R) \right] + F,
	\end{equation}
	\begin{equation}
		\begin{split}
			\tilde{Z} &= \varepsilon_m^2 (\epsilon_d^2 + \Gamma_L^2 - 2i \Gamma_L E) + 2 \Gamma_L \Gamma_R t_0^2  \\
			&-i E \left[ \Gamma_R (\epsilon_d^2 
			+\Gamma_L^2) + 2 t_0^2 (\Gamma_L + \Gamma_R) \right].
		\end{split}
	\end{equation}
\end{subequations}
The DOS in the dot is $\rho_{\rm{dot}} = -\frac{1}{2 \pi} \text{Im}\left[\mathcal{G}_e + \mathcal{G}_h \right]$.

To obtain the conductance, we first calculate the S-matrix using Eq. (\ref{smatrixGeneralEquation})
where $\mathcal{H}_{0}$ is given by Eq. (\ref{H0BdG}), and $W$ is

\begin{equation}
	W = \begin{pmatrix}t_1 & t_2 & 0 & 0 & 0 & 0 \\
		0 & 0 & 0 & -t_1 & -t_2 & 0 \\
		0 & 0 & 0 & 0 & 0 & 0 \\
		0 & 0 & t_R & 0 & 0 & -t_R \end{pmatrix},
\end{equation}
written in the basis $\{\Phi_{e, 1}, \Phi_{e, 2}, \Phi_{e, R}, \Phi_{h, 1}, \Phi_{h, 2}, \Phi_{h, R} \}$. Also, we must consider in Eq. (\ref{IL}) additional terms to account for the transmission coefficients to the third lead. However, in the limiting cases of interest, and, more generally, in the regime where $\varepsilon_m << \Gamma_{1, R}$ which we consider, the contributions of $T_{R1}$ and $A_{R1}$ are negligible and the conductance takes the same form of Eq. (\ref{GL}).

Noting that $\Sigma = -i \pi \rho W W^\dagger $, the matrix $(E - \mathcal{H}_{0} + i \pi \rho W W^\dagger)^{-1}$ is exactly the GF we calculated previously. To obtain the conductance, we use the probability of an incoming electron to be reflected as a hole in the same lead, $a_1$, and transmitted as an electron to leads $2$ and $R$, $t_{21}$ and $t_{R1}$, respectively, which we show below.

\begin{equation}
	t_{21} = -i\Gamma_L \mathcal{G}_e, \,\,t_{31} = -i\sqrt{\Gamma_L \Gamma_R} \Tilde{F}_{21}, \,\, a_1 = -i\Gamma_L F,
\end{equation}
where $\Tilde{F}_{21}$, in the low energy limit, is

\begin{equation}
	\tilde{F}_{21} \approx -i \varepsilon_m t_0 (\epsilon_d + i \Gamma_L + E ) \tilde{Z}^{-1}.
\end{equation}
The parameter regime for the analytical expressions in the main text consider $\epsilon_d >> \Gamma_L \sim t_0 >> \varepsilon_m$, and $E \to 0$. Besides, for $\Gamma_R = 0$, $t_{31} = 0$, $F \to 0$, suppressing Andreev reflection, and $T_{12} \approx \frac{\Gamma_L^2}{\epsilon_d^2 + \Gamma_L^2} \approx 0$. Therefore, $G (\Gamma_R = 0) \approx 0$. On the other hand, for $\Gamma_R >> \varepsilon_m^2\epsilon_d^2/t_0^2$,

\begin{equation}
	T_{31} = |t_{31}|^2 \approx  \frac{\varepsilon_m^2  \epsilon_d^2}{2 \Gamma_L \Gamma_R t_0^2} \approx 0,
\end{equation}
which is, as anticipated, again vanishingly small. Also, $\mathcal{G}_e \left(\Gamma_R >> \varepsilon_m^2\epsilon_d^2/t_0^2 \right) \approx F \left(\Gamma_R >> \varepsilon_m^2\epsilon_d^2/t_0^2 \right) \approx -i/2\Gamma_L$. Then, $T_{21} = A_1 = 1/4$. Finally,

\begin{equation}
	G \left(\Gamma_R >> \varepsilon_m^2\epsilon_d^2/t_0^2 \right) = \frac{e^2}{h}(A_1 + T_{21}) = \frac{e^2}{2h},
\end{equation}
which is precisely the result obtained for the infinite wire ($\varepsilon_m = 0$).

\section{Effective model -- temperature effects} \label{effectiveModelAndTemp}

In this section, we elaborate on the analytical results obtained for the effective model, $H_M = i\varepsilon_m\gamma_1 \gamma_2$, in Sec.~\ref{Majoranacond}. We first calculate the LDOS, $\rho_j (E) = -\dfrac{1}{\pi} {\rm Im}[\mathcal{G}_{jj}^r(E)]$, which in this simple model has only two sites: left $L$ and right $R$. The retarded Green function $\mathcal{G}^r(E)$ is given by 
\begin{equation}
	\mathcal{G}^r(E) = \left(E - H_{M} -\Sigma\right)^{-1},
\end{equation}
where $\Sigma=-i\pi \rho W W^\dagger$ is the self-energy and $W$ describes the coupling between the Majorana modes and the left/right metallic leads~\cite{splittingBeenakker},
\begin{equation}
	W = \begin{pmatrix}
		t_L & 0 & t_L^* & 0 \\
		0 & t_R &  0 & t_R^*
		\end{pmatrix}.
\end{equation}
Using
\begin{equation}
	\Sigma = \begin{pmatrix}
		-i\Gamma_L & 0 \\
		0 & -i\Gamma_R
	\end{pmatrix},
\end{equation}
with $\Gamma_\alpha = 2\pi\rho|t_\alpha|^2$, we then obtain
\begin{equation}
	\mathcal{G}^r(E) =\dfrac{1}{(E+i\Gamma_L)(E+i\Gamma_L)-\varepsilon_m^2} \begin{pmatrix}
		E+i\Gamma_R & i\varepsilon_m \\
		-i \varepsilon_m & E+i\Gamma_L
	\end{pmatrix}.
	\label{GF}
\end{equation}
The LDOS extracted from \eqref{GF} reads
\begin{equation}
	\rho_\alpha(E) = \dfrac{1}{\pi} \dfrac{(E^2+\Gamma_\beta^2)\Gamma_\alpha + \varepsilon_m^2\Gamma_\beta}{\varepsilon_m^2[\varepsilon_m^2 - 2(E^2-\Gamma_\alpha \Gamma_\beta)] + (E^2+\Gamma_\alpha^2)(E^2+\Gamma_\beta^2) }.
	\label{rhoE}
\end{equation}
with $\alpha = L ~(R)$ and $\beta=R~(L)$.

\begin{figure}[htb!]
	\centering
	\includegraphics[width=0.235\textwidth, keepaspectratio]{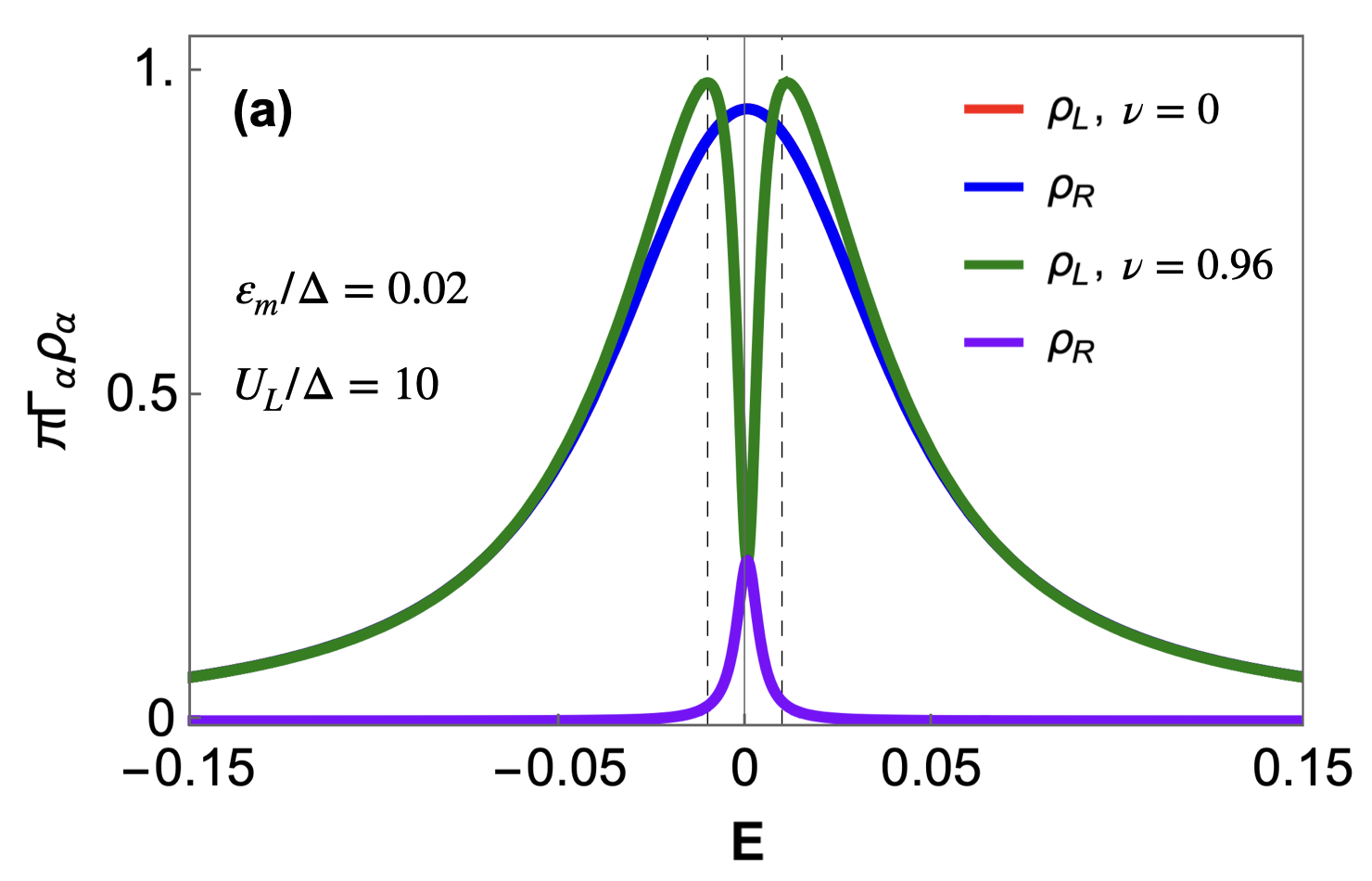}
	\centering
	\includegraphics[width=0.235\textwidth]{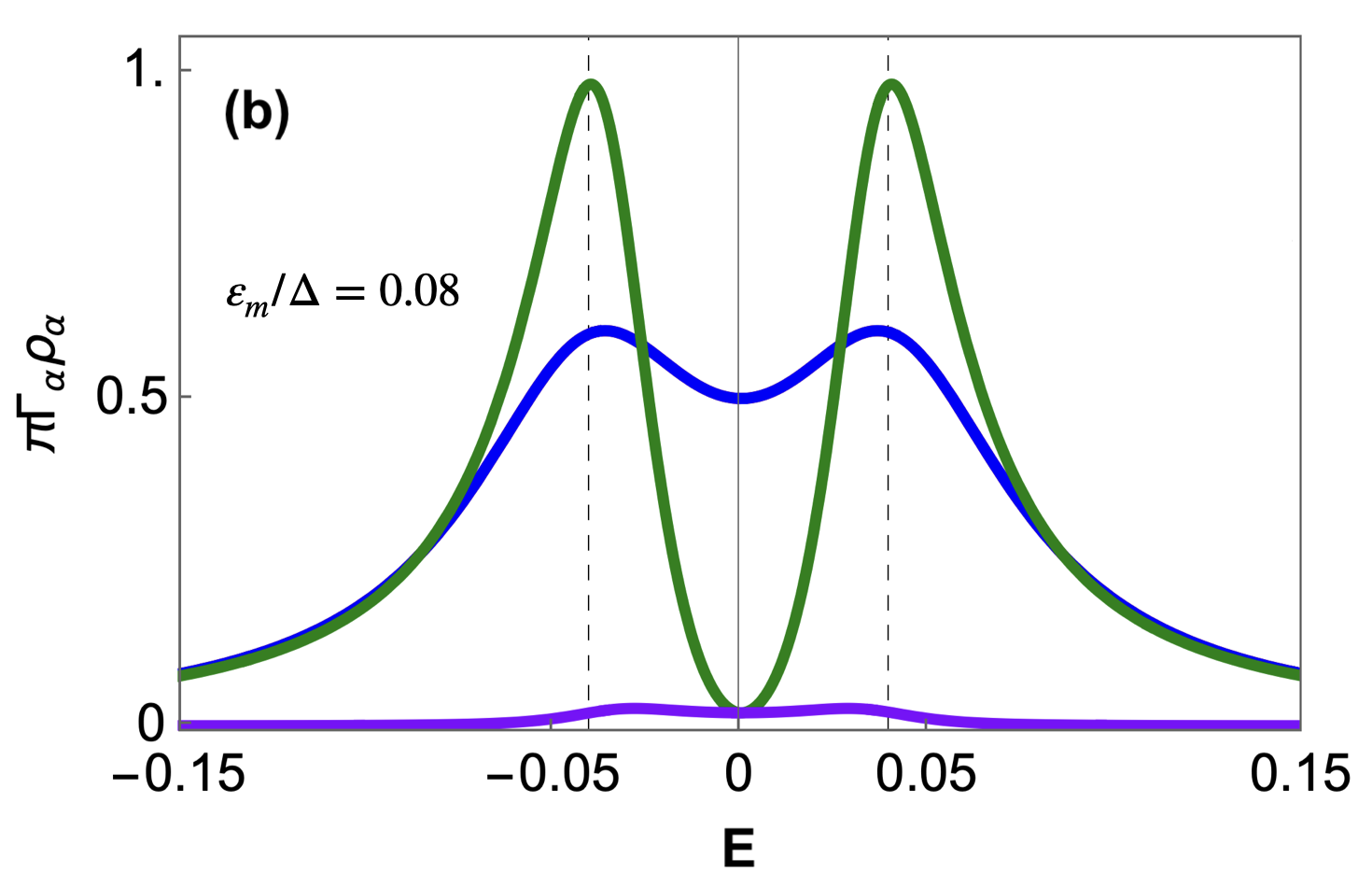}
	\caption{Local density of states $\rho_L$ and $\rho_R$ as functions of $E$ for (a) $\varepsilon_m/\Delta=0.02$ and (b) $\varepsilon_m/\Delta=0.08$ in the symmetric ($\Gamma_L=\Gamma_R$, $\nu=0$) and asymmetric  ($\Gamma_R=0.02\Gamma_L$, $\nu=0.96$) configurations. }
	\label{ldosxE}
\end{figure}

In Fig.~\ref{ldosxE}, we show $\rho_L$ and $\rho_R$ as functions of $E$ for two different values of the hybridization energy $\varepsilon_m$. In this effective model, $\varepsilon_m$ controls the wire length $L_s$, i.e., the larger $\varepsilon_m$, the shorter the wire. In the full Hamiltonian, $\varepsilon_m$ is not only connected to $L_s$, but also to other physical parameters like the Zeeman energy and the superconducting correlation lentgh~\cite{DasSarmaSplitting}. We choose parameters consistent with the full Hamiltonian presented in the main text.  The goal here is to emphasize the crucial role played by $\varepsilon_m$ in our main result: the nonlocality of the local conductances.  

We start by analyzing $\varepsilon_m=0$, which yields the usual Lorentzian-shape LDOS,
\begin{align}
	\rho_\alpha=\dfrac{1}{\pi}\dfrac{\Gamma_\alpha}{E^2+\Gamma_\alpha^2}.
	\label{rho0}
\end{align}
According to our proposal (see Fig. 3 of the main text), we now vary, for example, the coupling $\Gamma_R$ and keep $\Gamma_L$ fixed. It is clear from Eq.~\eqref{rho0} that $\rho_L$ does not change, i.e, variations on the right side do not affect the left. Let us then turn on the hybridization energy. In Figs.~\ref{ldosxE} (a) and (b),  we take $\varepsilon_m=0.01$ and $0.04$ meV, respectively, and show $\rho_L$ and $\rho_R$ for the symmetric ($\Gamma_L=\Gamma_R$) and asymmetric ($\Gamma_R=0.02\Gamma_L$) cases. We observe that in the symmetric configuration, $\rho_L(E)=\rho_R(E)$, red and blue curves, for any magnitude of the hybridization energy. When $\Gamma_R \ne \Gamma_L$, on the other hand, $\rho_L$ (green curve) has a dip at $E=0$ and is maximum at $\pm \varepsilon_m$ (vertical dashed lines), while $\rho_R$ (purple curve) is enhanced around $E=0$ (not visible in the figures because we are plotting $\Gamma_R \rho_R$, with $\Gamma_R \ll 1$ for $\nu=0.96$). These features can be easily seen by taking $E=0$ in Eq.~\eqref{rhoE},
\begin{align}
	\rho_L(0) &=\dfrac{1}{\pi}\dfrac{\Gamma_R}{\varepsilon_m^2+\Gamma_L \Gamma_R}, \\
	\rho_R(0) &=\dfrac{1}{\pi}\dfrac{\Gamma_L}{\varepsilon_m^2+\Gamma_L \Gamma_R}.
\end{align}
We note that, given  $\Gamma_L$ and $\Gamma_R$, the dip and its width are controlled by $\varepsilon_m$. The results above are in agreement with Figs. 3 (a) and (b) of the main text. Next, we use the LDOS to analyze temperature effects on the conductance.

\begin{figure}[htb!]
	\centering
	\includegraphics[width=0.35\textwidth, keepaspectratio]{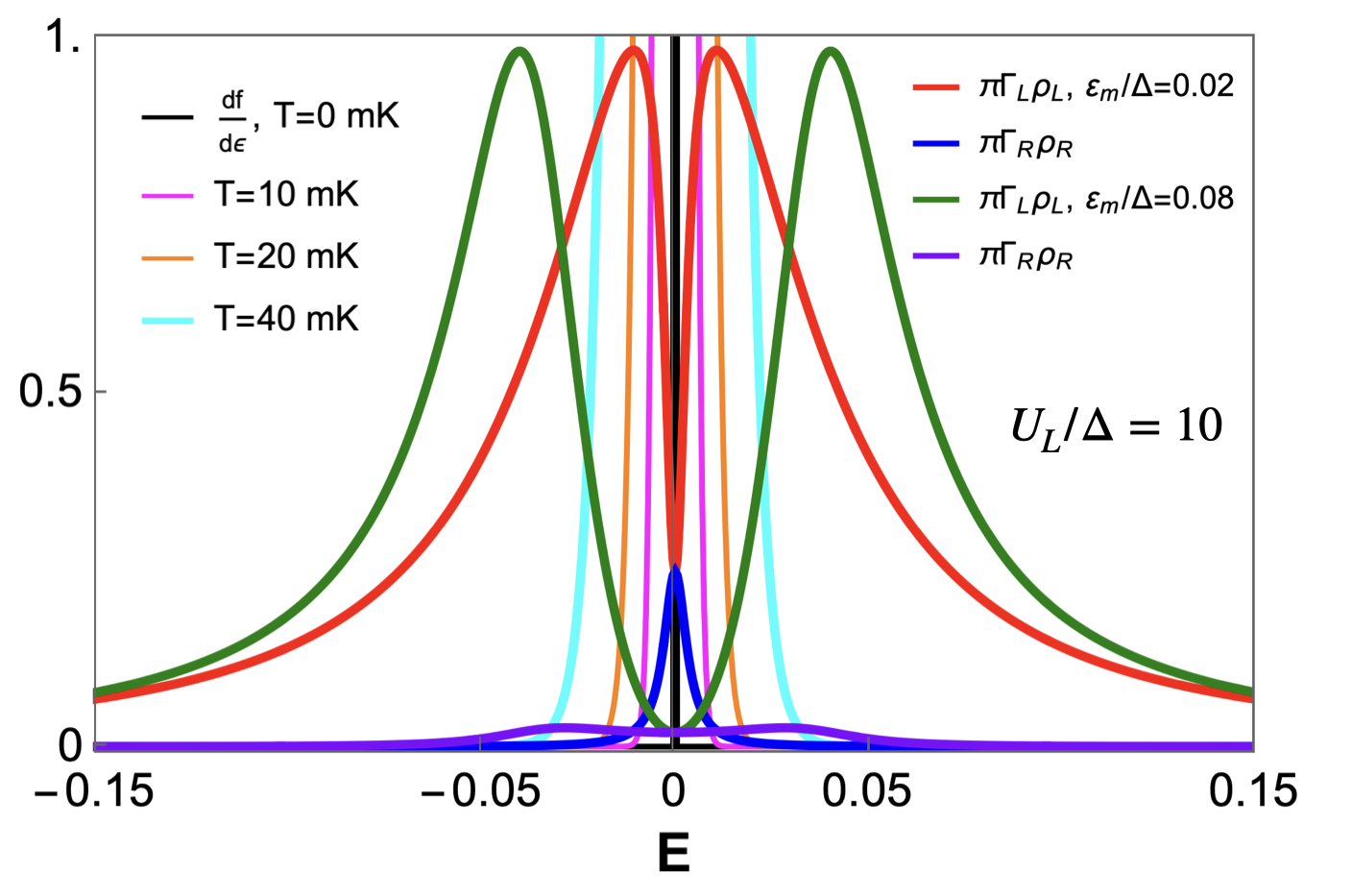}
	\caption{Local density of states $\rho_L$, $\rho_R$, and derivative of the Fermi function $df/dE$ vs. $E$, for different tempeartures $T=0, 10, 20, 40$~mK and two magnitudes of $\varepsilon_m$, with $\nu=0.96$.}
	\label{LDOSxT}
\end{figure}

The local conductance $G_{\alpha \alpha}$ at $T \ne 0$ and small bias voltage is given by 
\begin{equation} \label{G_aa}
	G_{\alpha \alpha}  = \dfrac{e^2}{h}\int dE{\left[2 A_\alpha(E) + T_{\beta \alpha}(E) + A_{\beta \alpha}(E) \right] \frac{df(E)}{dE}},
\end{equation}
which can be rewritten as
\begin{equation}
	G_{\alpha \alpha}  = \dfrac{ e^2}{h}\int dE{\left[2\pi \Gamma_\alpha \rho_\alpha(E) \right]\frac{df(E)}{dE}},  
	\label{Grho}
\end{equation}
with $df/dE$ the derivative of the Fermi function. In Fig.~\ref{LDOSxT} we show $\rho_L$, $\rho_R$, and $df/dE$ as functions of $E$ for different temperatures $T=0, 10, 20, 40$~mK and $\varepsilon_m/\Delta=0.02, 0.08$ in the asymmetric configuration ($\nu=0.96$). 

\begin{figure}[htb!]
	\centering
	\includegraphics[width=0.235\textwidth, keepaspectratio]{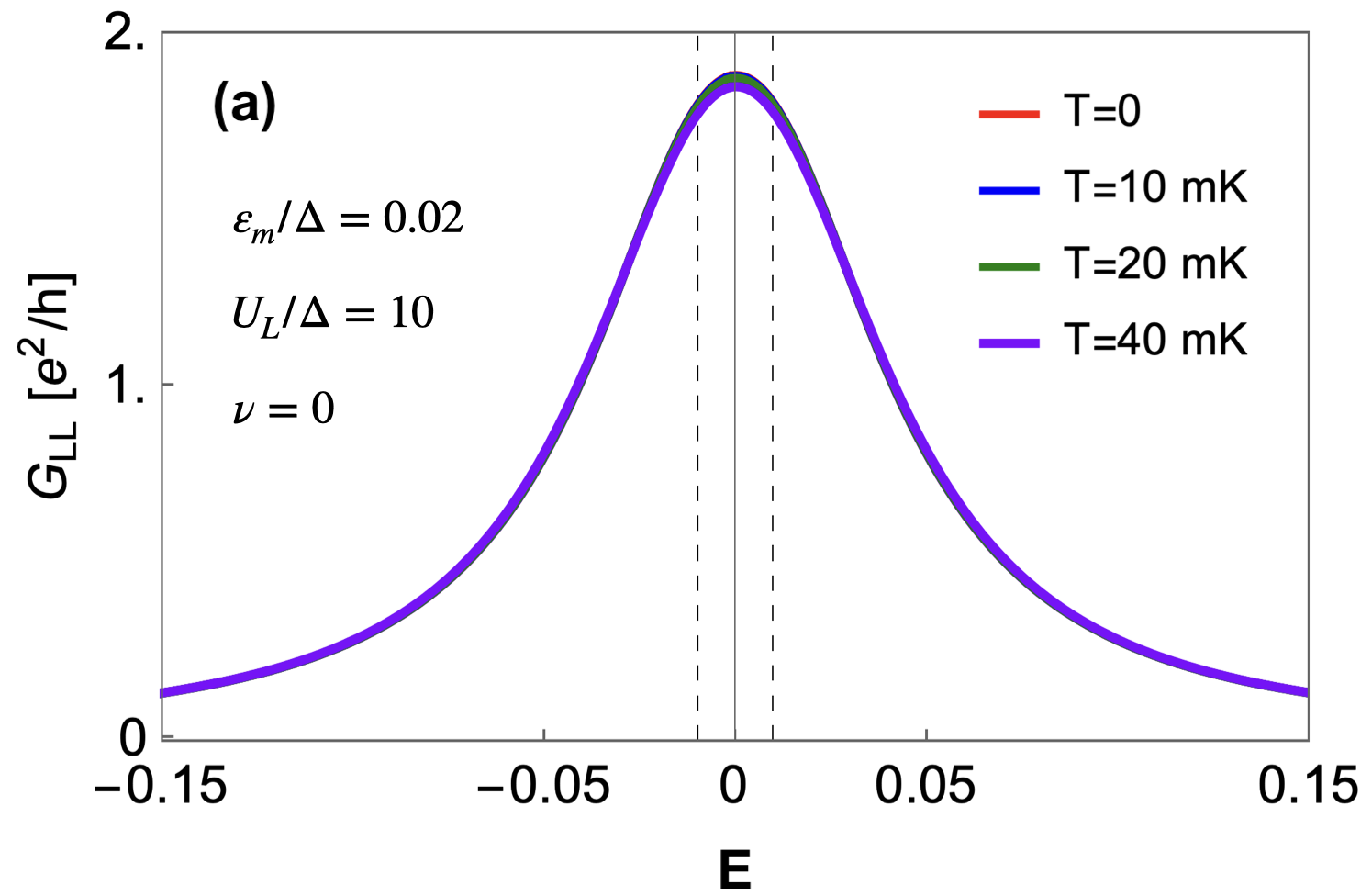}
	\centering
	\includegraphics[width=0.235\textwidth]{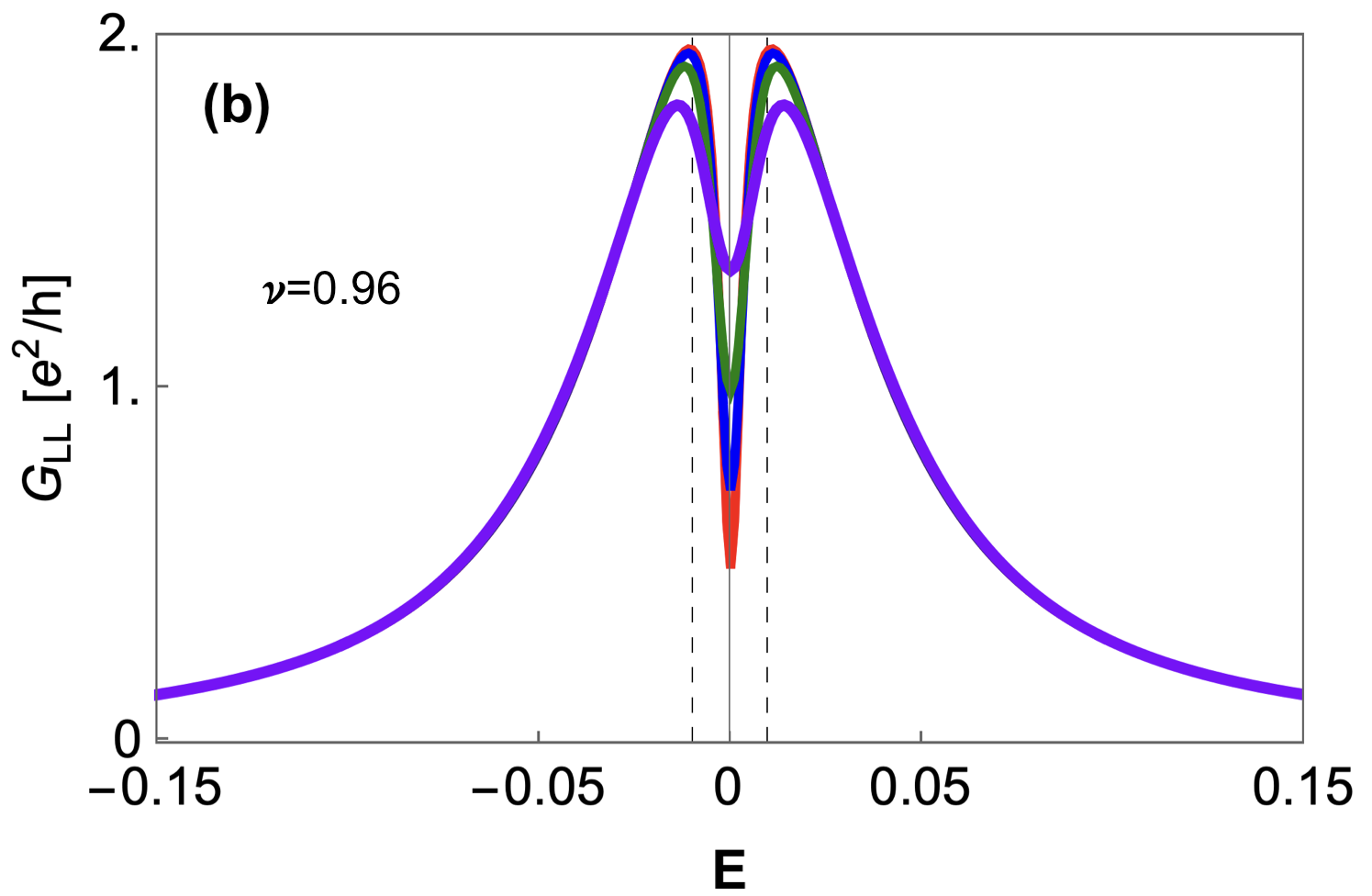}
		\centering
	\includegraphics[width=0.235\textwidth, keepaspectratio]{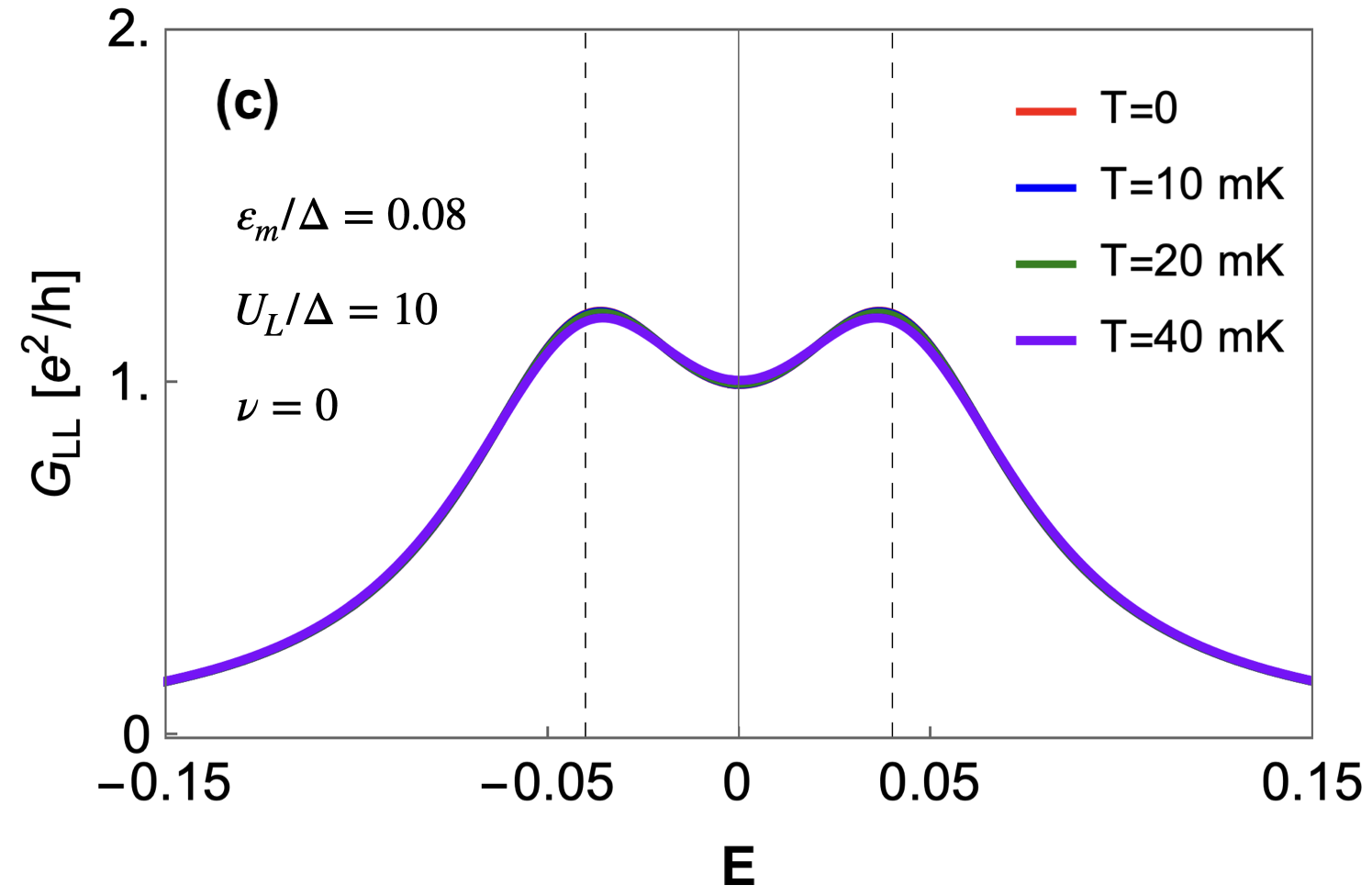}
	\centering
	\includegraphics[width=0.235\textwidth]{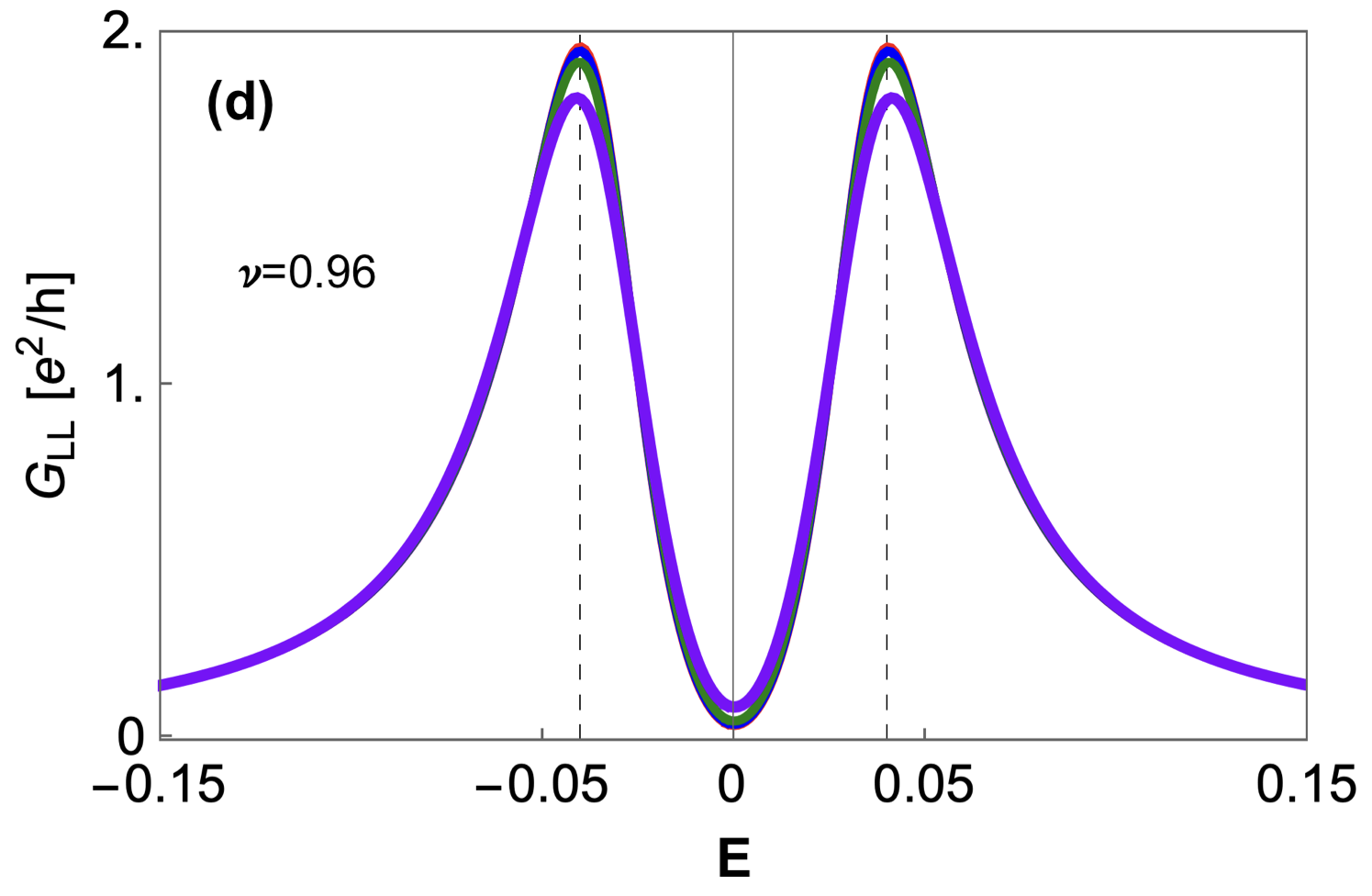}
	\caption{Conductance $G_{LL}$ as a function of $E$ for $\varepsilon_m/\Delta=0.02$, (a) $\nu=0$, and (b) $\nu=0.96$, and $\varepsilon_m/\Delta=0.08$, (c) $\nu=0$, and (d) $\nu=0.96$ for $T=0, 10, 20, 40$~mK.}
	\label{GxE}
\end{figure}

When we increase $T$ (see, for instance, the orange and cyan curves), values of $\rho_\alpha$ away from $E=0$ start to contribute more, increasing the value of the conductance $G_{\alpha \alpha}$. As discussed above, in the asymmetric configuration, $\varepsilon_m$ controls the spectral weight at $E=0$, especially on the left side of the wire (red and green curves). Hence when $\varepsilon_m$ is such that the width of the dip in $\rho_L$ is wider than the temperature broadening, i.e., $\varepsilon_m \gg k_BT$, the effect of $T$ on the conductance is negligible. This can be seen in Fig.~\ref{GxE}, where we show $G_{LL}$ as a function of $E$ for $\varepsilon_m/\Delta=0.02$, (a) and (b), and $\varepsilon_m/\Delta=0.08$, (c) and (d). For comparison, in (a) and (c), we show the symmetric configuration. 

\begin{figure}[htb!]
	\centering
	\includegraphics[width=0.235\textwidth, keepaspectratio]{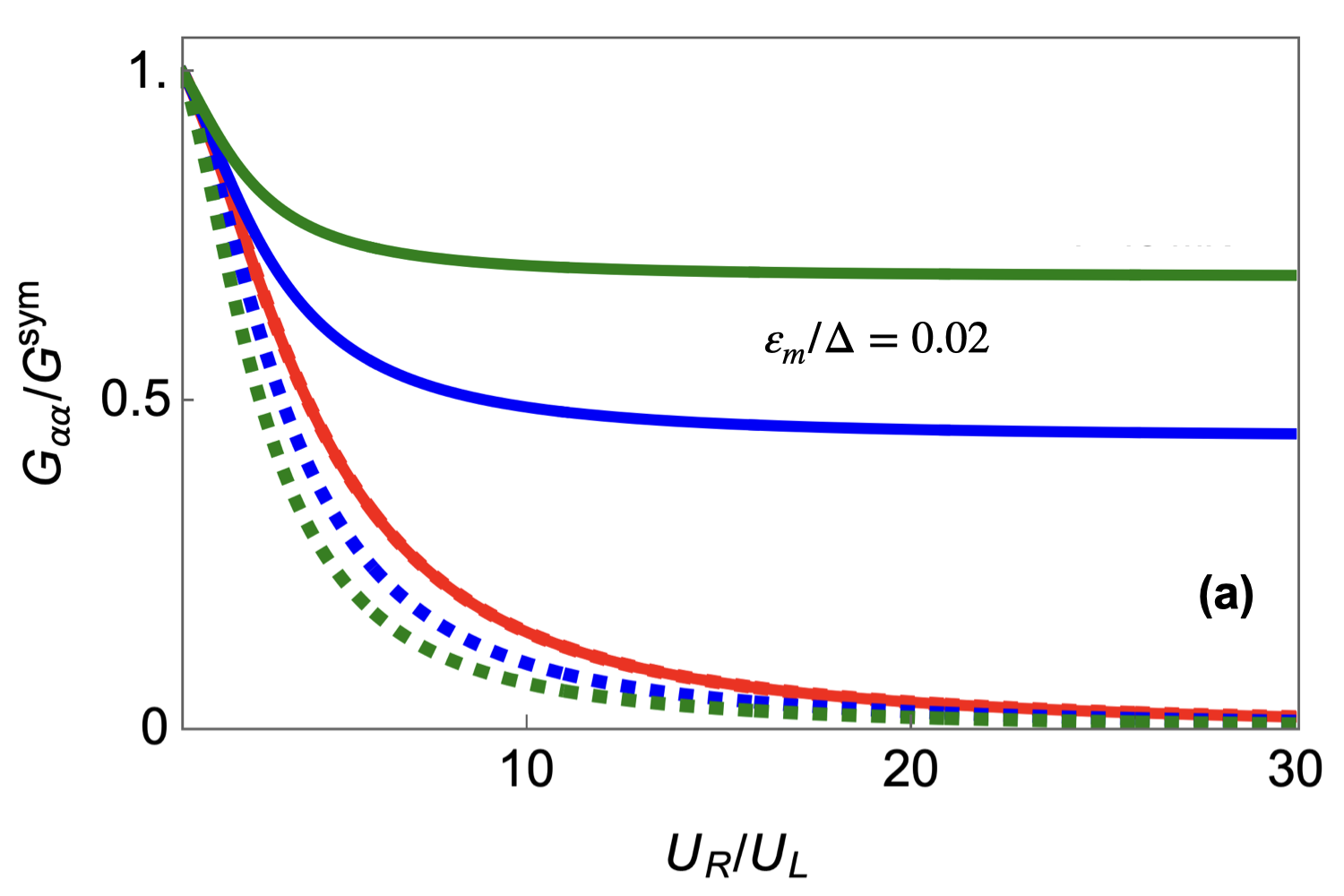}
	\centering
	\includegraphics[width=0.235\textwidth]{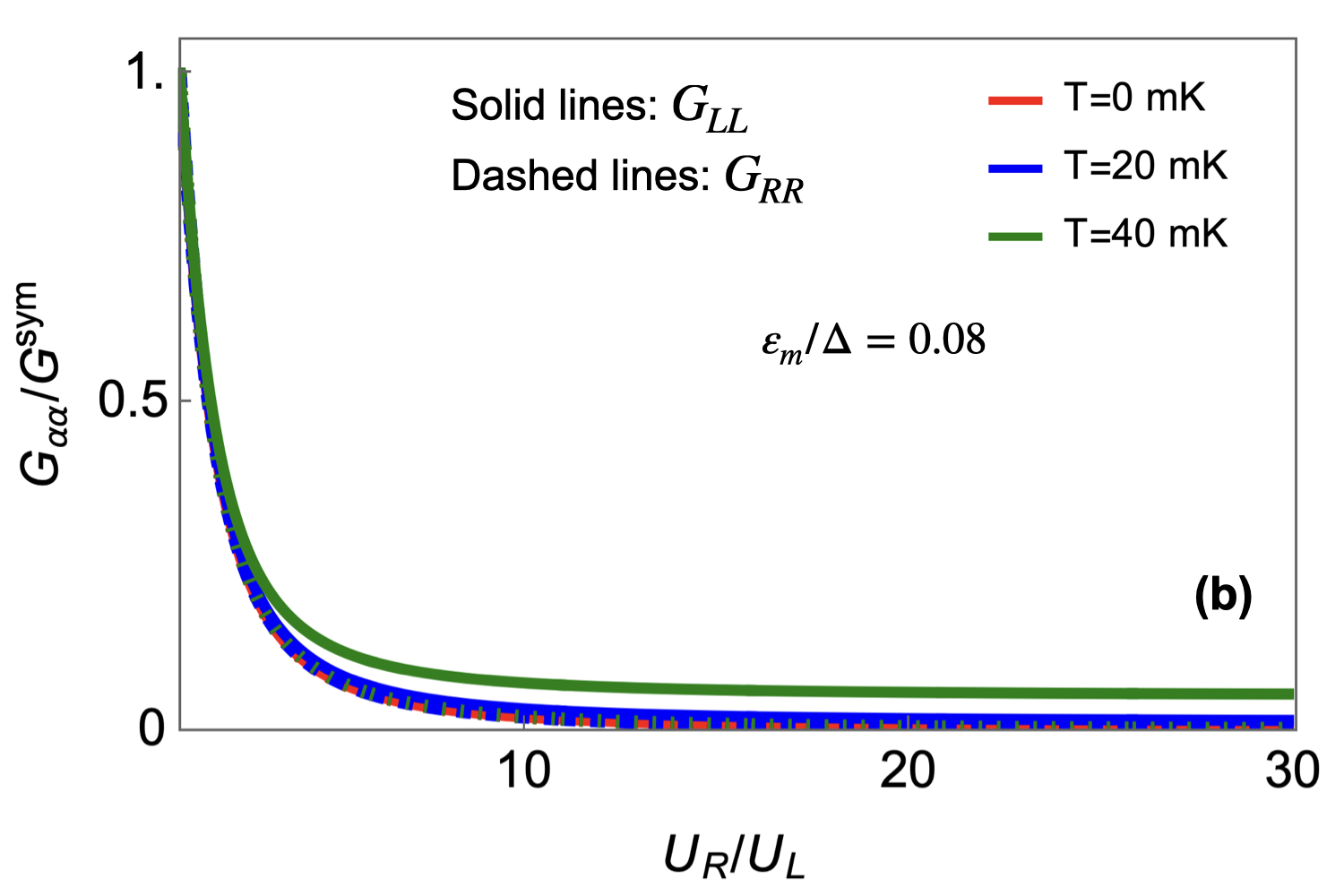}
	\caption{Conductances $G_{LL}$ (solid lines) and $G_{RR}$ (dashed lines) as functions of $U_R/U_L$ for (a) $\varepsilon_m/\Delta=0.02$ and (b) $\varepsilon_m/\Delta=0.08$ for $T=0, 20, 40$~mK.}
	\label{condxU}
\end{figure}

In Fig.~\ref{condxU}, we  show the dependence of $G_{LL}$ (solid lines) and $G_{RR}$ (dashed lines) as we vary $U_R/U_L$ for nonzero $T$. Here we assume $\Gamma_\alpha \propto 1/U_\alpha^2$. At $T=0$, $G_{LL}=G_{RR}$, as predicted by the full model. As $T$ increases, as long as $\varepsilon_m \gg k_BT$ [as in (b)],  $G_{LL} \approx G_{RR}$, in agreement with the results in Fig. 5(a) of the main text.

\section{Local conductance oscillations as a function of the Zeeman field} \label{GLLoscillationsWithVz}

\begin{figure}[htb!]
\centering
\includegraphics[width=0.235\textwidth]{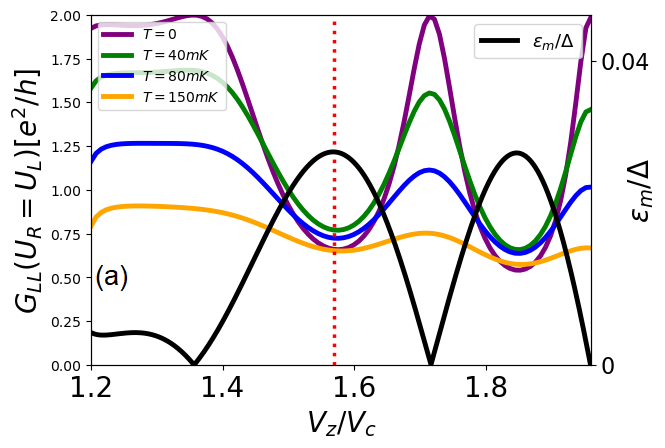}
\centering
\includegraphics[width=0.235\textwidth]{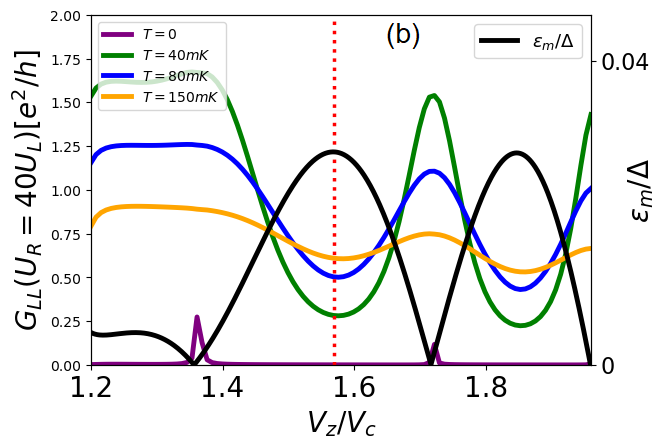}
\caption{(a) $G_{LL}(U_L = U_R)$ as a function of $V_z$. (b) $G_{LL}(U_R = 40U_L)$ as a function of $V_z$. The combination of the two plots results in the Majorana oscillations shown in Fig. 5(b) of the main text.}
\label{GLL oscillations}
\end{figure}

\begin{figure}[htb!]
	\centering
	\includegraphics[width=0.32\textwidth]{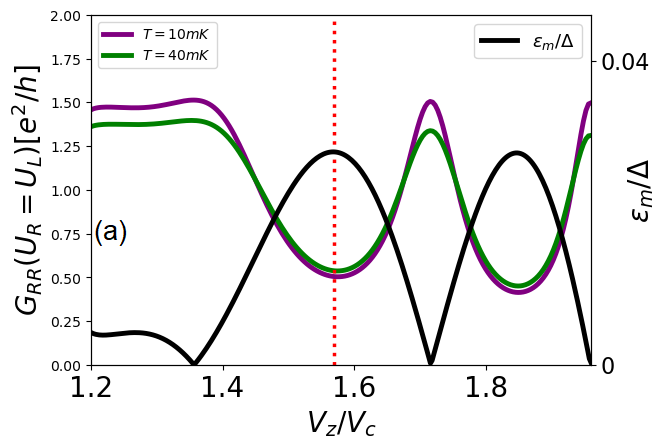}
	\centering
	\includegraphics[width=0.32\textwidth]{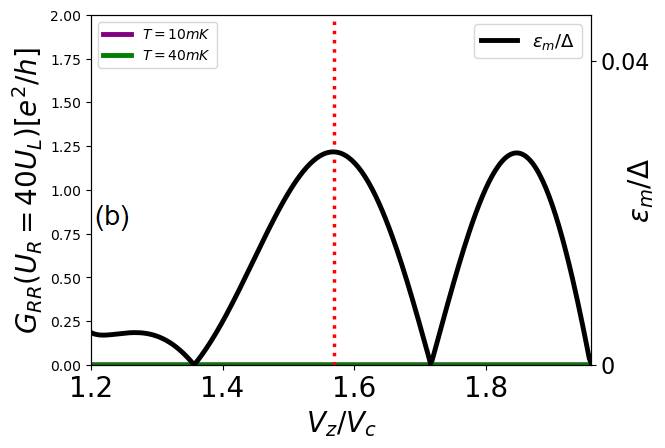}
	\centering
	\includegraphics[width=0.32\textwidth]{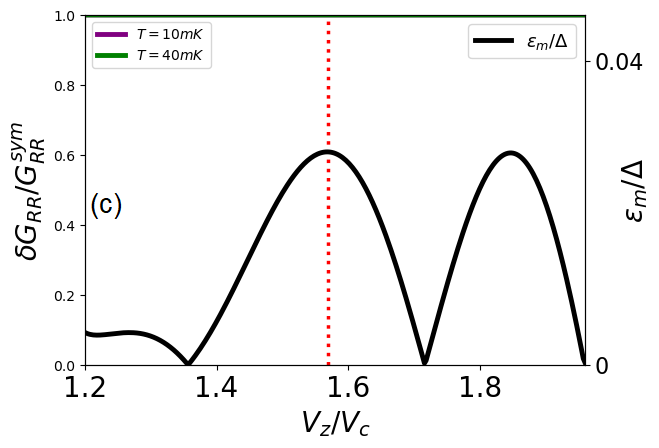}
	\caption{(a) $G_{RR}(U_L = U_R)$, (b) $G_{RR}(U_R = 40U_L)$, and (c) $\delta G_{RR}$ as functions of $V_z$.}
	\label{GRR oscillations}
\end{figure}

In the main text, we have shown how the asymmetric conductance deviation $\delta G_{LL}^{sym} = G_{LL}^{U_R = U_L} - G_{LL}^{U_R \gg U_L}$ allows one to perfectly track Majorana oscillations in the regime $\varepsilon_m/k_B T << 1$. For completeness, here we  show how the local conductance oscillates in both cases, i.e., $U_L = U_R$ and $U_L << U_R$, Fig. \ref{GLL oscillations}. In addition, we show the dependence of $G_{RR}$ as a function of $V_z/V_c$, in the symmetric and asymmetric cases, Fig.~\ref{GRR oscillations} (a) and (b), respectively, and $\delta G_{RR}$, Fig.~\ref{GRR oscillations} (c).

\section{Finite bias voltages} \label{FiniteBiasSec}

In this section, we study the effect of a nonzero bias voltage in our conductance simulations. At finite bias voltages, the density of states, see Fig. 3(a) of the main text, is scanned via the term $df(E - eV_L)/dE$ in Eq. (\ref{G_aa}), which selects the range of energies that contribute to the conductance.

\begin{figure}[htb!]
	\centering
	\includegraphics[width=0.235\textwidth, keepaspectratio]{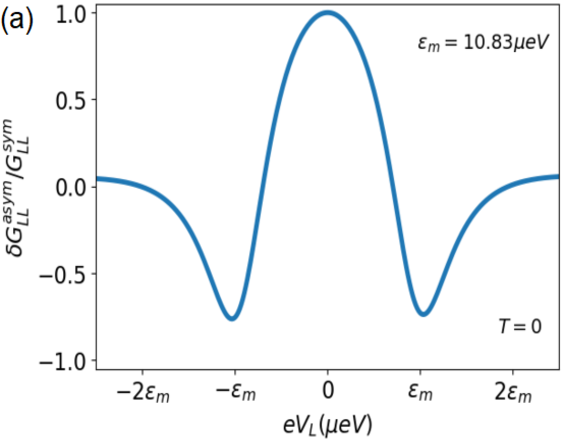}
	\centering
	\includegraphics[width=0.235\textwidth]{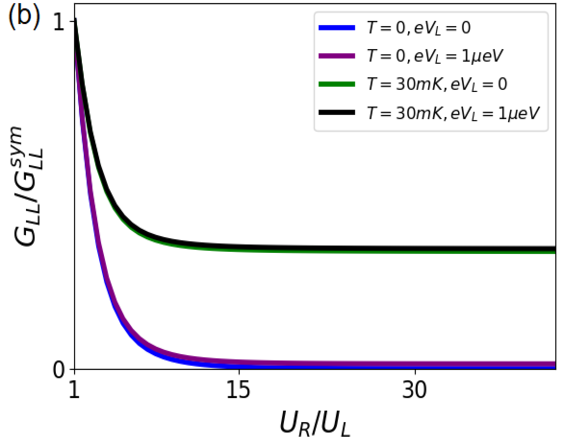}
		\centering
	\includegraphics[width=0.235\textwidth, keepaspectratio]{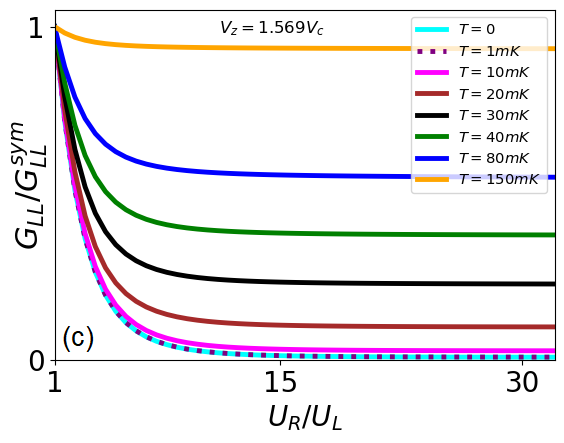}
	\centering
	\includegraphics[width=0.235\textwidth]{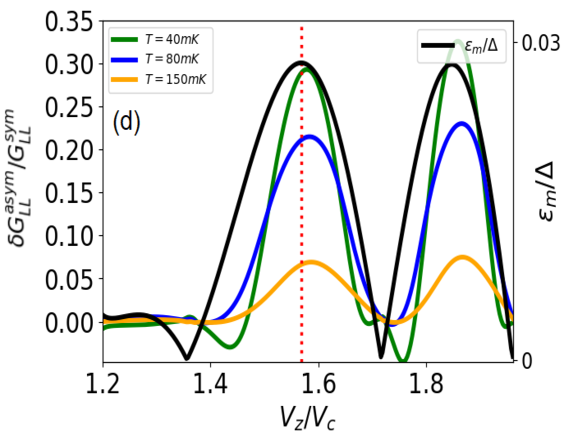}
	\caption{Conductance simulations considering finite bias voltages. (a) $\delta G_{LL}^{asym}/G_{LL}^{sym}$ as a function of the bias voltage for the pristine wire shown in Fig. \ref{fig1}(a) of the main text. The hybridization energy for this case is $\varepsilon_m = 10.83~\mu$eV. (b) $G_{LL}$ vs. $U_R/U_L$ for $V_L = 0$, and $1~\mu$V, for $T = 0$, blue and purple curves, respectively, and also for $T = 30$~mK, green and black curves, respectively. (c)-(d) We repeat the simulations for Fig. 5 of the main text with the same parameters but considering $V_L = 1~\mu$V.}
	\label{finiteBias}
\end{figure}

We first analyze the pristine case considered in Fig. 1 of the main text ($L_s = 2~\mu $m and $V_z = 1.65 V_c$). In Fig. \ref{finiteBias} (a), we show $\delta G_{LL}^{asym}$ as a function of $eV_L$. We observe that $\delta G_{LL}^{asym}$ is maximized at $V_L = 0$ and is  suppressed as $V_L$ increases. This occurs because in the highly asymmetric situation ($\nu \to 1$) the minimum value of the LDOS is at $E = 0$, which causes the $G_{LL}$ suppression. Therefore, if we move away from zero bias, the visibility of the nonlocality we want to show via $G_{LL}(U_R \gg U_L)$ is reduced. Interestingly, at $eV_L = \pm \varepsilon_m$, there are dips in $\delta G_{LL}^{asym}$. This effect arises from an increase in the LDOS at $E = \pm \varepsilon_m$ as $\Gamma_R$ decreases, see Fig. 3 (a) of the main text. Hence at $eV_L = \pm \varepsilon_m$, $G_{LL}$ has a maximum value for $U_R \gg U_L$, which, in turn, causes the dips in $\delta G_{LL}^{asym}$.

Although the suppression of $G_{LL}$ upon the increase of $U_R$ is maximum for $V_L = 0$, one must account for errors in setting the zero-bias conductance measurements. In experiments, see for instance Ref.~\cite{heedt2021shadow}, the linewidth of zero-bias conductance peaks is, in general, $< 10~\mu$eV. Considering the error to be of the order of $1~\mu$V, i.e. $V_L = 0 \pm 1 ~\mu $V, we simulate the results in the main text for the maximum deviation, $V_L = 1~\mu$V. In Fig. \ref{finiteBias}(b), we show that for the pristine wire in Fig. 1 (b) of the main text, the error in the bias voltage does not alter the results significantly. This is observed for $T = 0$, blue and purple curves, and also for $T = 30 $~mK, black and green curves. Similarly, we perform simulations for the disordered wire shown in Fig. 5 using $V_L = 1~\mu$V. The results are shown in Figs. \ref{finiteBias} (b) and (c). We note that the results presented in the main text remain qualitatively the same, with a small shift on the peak of $\delta G_{LL}^{asym}$ oscillations. We conclude, therefore, that our results are robust against deviations in the bias voltage on zero-bias conductance measurements.  

\newpage
\bibliography{Refs}

\end{document}